\documentclass[preprint2,twocolumn,times,tighten]{aastex63}

\usepackage{color}
\usepackage{enumitem}
\usepackage{hyperref}
\usepackage{verbatim}
\usepackage{amsmath,amstext,mathrsfs}
\usepackage[all]{hypcap} 
\usepackage{afterpage}    
\usepackage{marginnote}
\usepackage{makecell}
\usepackage{subfigure}
\usepackage{multirow}

\shorttitle{Magnetic Field Morphology in Galaxy Clusters with the Gradient Technique}
\shortauthors{Hu et al.}
\graphicspath{{./}{figures/}}

\begin{document}
	
	\title{Probing Magnetic Field Morphology in Galaxy Clusters with the Gradient Technique}

	\email{yue.hu@wisc.edu, alazarian@facstaff.wisc.edu}
	
	\author[0000-0002-8455-0805]{Yue Hu}
	\affiliation{Department of Physics, University of Wisconsin-Madison, Madison, WI 53706, USA}
	\affiliation{Department of Astronomy, University of Wisconsin-Madison, Madison, WI 53706, USA}
	\author{A. Lazarian}
	\affiliation{Department of Astronomy, University of Wisconsin-Madison, Madison, WI 53706, USA}
	\affiliation{Center for Computation Astrophysics, Flatiron Institute, 162 5th Ave, New York, NY 10010}
	\author{Yuan Li}
	\affiliation{Department of Astronomy, University of California, Berkeley, CA 94720, USA}
	\author{Irina Zhuravleva}
	\affiliation{Department of Astronomy \& Astrophysics, University of Chicago, Chicago, IL 60637, USA}
	\author{Marie-Lou Gendron-Marsolais}
	\affiliation{European Southern Observatory, Alonso de Cordova 3107, Vitacura, Casilla 19001, Santiago, Chile}
	
	\begin{abstract}
		Magnetic fields in the intracluster medium (ICM) affect the structure and the evolution of galaxy clusters. However, their properties are largely unknown, and measuring magnetic fields in galaxy clusters is challenging, especially on large-scales outside of individual radio sources.
		In this work, we probe the plane-of-the-sky orientation of magnetic fields in clusters using the intensity gradients. The technique is a branch of the Gradient Technique (GT) that employs emission intensity maps from turbulent gas. We utilize the {\it Chandra} X-ray images of the Perseus, M 87, Coma, and A2597 galaxy clusters, and the VLA radio observations of the synchrotron emission from Perseus. We find that the fields predominantly follow the sloshing arms in Perseus, which is in agreement with numerical simulations. The GT-predicted magnetic field shows signatures of magnetic draping around rising bubbles driven by supermassive black hole (SMBH) feedback in the centers of cool-core clusters, as well as draping around substructures merging with the Coma cluster. We calculate the mean-field orientation with respect to the radial direction in these clusters. In the central regions of cool-core clusters, the mean orientation of the magnetic fields is preferentially azimuthal. There is a broad agreement between the magnetic field of Perseus predicted using the X-ray and radio data.  Further numerical studies and better future observations with higher resolution and the larger effective area will help reduce the uncertainties of this method.
		
	\end{abstract}
	

	\section{Introduction}

	\label{sec:intro}
	Magnetic fields are pervading in multi-scale astrophysical environments, including protostars, molecular clouds, galaxies and galaxy clusters \citep{1995ApJ...443..209A,2002RvMP...74..775W,2010ApJ...710..853C,2012ARAA...50..29,2006ApJ...647..374G,1994RPPh...57..325K}. They play a crucial role in regulating accretion flows, transport processes (e.g., transport of heat), and cosmic rays propagation. The primary ways of observing cosmic magnetic fields include the measurement of synchrotron emission, Faraday rotation, and Zeeman splitting. Synchrotron radiation arises from relativistic electrons spiraling along magnetic field lines \citep{2006ApJS..167..230H,2006AJ....131.2900C,2019ApJ...871..106D}, and can be used to estimate the strength and the orientation of the magnetic fields in the plane-of-the-sky (POS)  \citep{2011MNRAS.412.2396F,2012A&ARv..20...54F}. Faraday rotation and Zeeman measurements are used to measure the magnetic fields along the line-of-sight (LOS, \citealt{2010A&A...513A..30B, 2015A&A...575A.118O,2016ApJ...830...38L,2012ARAA...50..29,2019FrASS...6...66C}).
	
	Galaxy clusters are the largest non-linear systems in the universe, filled with weakly-magnetized intracluster medium \citep[e.g.,][]{2008SSRv..134...93F,2002ARA&A..40..319C,2012A&ARv..20...54F}. 
	The weak magnetic fields in clusters are likely unimportant dynamically on large scales but can change the microscopic properties of the plasma \citep[e.g.,][]{2005ApJ...629..139S,2014PhRvL.112t5003K}. For instance, numerical simulations and observations of gas sloshing in galaxy clusters show that magnetic fields can suppress the development of the  Kelvin-Helmholtz instabilities, mixing of two gas phases, and transport processes \citep[see][for reviews]{2007PhR...443....1M,2016JPlPh..82c5301Z}. Observational evidence of magnetic fields modifying bulk plasma properties on microphysical scales has recently been found in the Coma cluster \citep{IZ19}. Observed structure functions of turbulence in the cold filaments in the innermost regions of galaxy clusters are consistent with this conclusion \citet{2019arXiv191106329L}.
	
	Despite the importance of magnetic fields in galaxy clusters, probing their properties with observations remains challenging. Faraday rotation measures the magnetic fields along the LOS, and has shown that the fields in clusters are weak with strength in the order of a few $\mu$G \citep[e.g.,][]{2003A&A...412..373V, 2010A&A...513A..30B}. The low magnetic field strength limits the applicability of Zeeman measurements in clusters. Also, there is a lack of observational constraints on the POS magnetic field component \citep{1996ARA&A..34..155B,2019arXiv191111163P,2013Natur.501..391E}.
	
	Clusters are dynamical objects that accrete matter along the cosmic-web filaments, undergo many minor and major mergers, and host feedback processes from central active galactic nuclei (AGN). Therefore, one expects that the ICM is in a turbulent state. Indeed, the {\it Hitomi} X-ray satellite measured velocities $\simeq 100-200$ km/s in the central region of the Perseus cluster \citep{2018PASJ...70...11H}; resonant scattering and Doppler broadening measurements revealed velocities $\sim 100$ km/s in the central region of groups and massive galaxies \citep[e.g.,][]{2009MNRAS.398...23W, 2017MNRAS.472.1659O, 2018PASJ...70...10H,2010MNRAS.402L..11S}; similar level of turbulence has been measured indirectly through the analysis of X-ray surface brightness fluctuations \citep[e.g.,][]{2004A&A...426..387S,2012MNRAS.421.1123C,2014Natur.515...85Z,2015MNRAS.453.3699W}.
	
	Rotation Measures suggest that the magnetic fields in the ICM are also turbulent \citep{2019MNRAS.487.4768S,2006A&A...453..447E,1993ApJ...411..518G}. 
	The turbulent fluctuations in both synchrotron radiation and X-ray emission can also provide information about the magnetic fields morphology, which is the scope of this work.
	
	We have developed an innovative Gradients Technique (GT) to study magnetic fields across multiple scales based on the anisotropic properties of MHD turbulence, i.e., turbulent eddies are elongated along the local magnetic fields \citep{GS95,LV99}. As a result, the density gradient and velocity gradient of the turbulent eddies are perpendicular to the magnetic fields \citep{CL02,2003MNRAS.345..325C,2002ApJ...564..291C}. This anisotropic behavior is also expected in the ICM (see \S~\ref{sec:theory} for a detailed discussion). Therefore, the magnetic field in clusters can be inferred from the density gradient of either synchrotron or X-ray radiation. \citet{2018ApJ...855...72L} shows the Synchrotron Intensity Gradients (SIGs) can be used to trace the magnetic field component lying on the POS, while the Synchrotron Polarization Gradients can measure the LOS magnetic field component (SPGs, see \citealt{2018ApJ...865...59L}). The density gradient calculated from spectroscopic data is denoted as the Intensity Gradients (IGs, see \cite{IGs}). When the velocity information is available,  the Velocity Centroid Gradients (VCGs, see \citealt{2017ApJ...835...41G}) and the Velocity Channel Gradients (VChGs, see \citealt{2018ApJ...865...46L}) can also be used to study magnetic field.

	The applicability of GT in tracing the magnetic field morphology has been numerically and observationally tested in many astrophysical environments, including protostars \citep{2019ApJ...880..148G}, molecular clouds \citep{velac,survey,2019ApJ...873...16H,2020arXiv200715344A}, galactic diffuse transparent gas \citep{PCA,2019ApJ...874...25G,EB,PDF, YL17a}, and synchrotron emissions from the ISM \citep{2018ApJ...855...72L,2018ApJ...865...59L,2019ApJ...887..258H,2019MNRAS.486.4813Z}. Several studies also extend the GT to estimate the magnetization level \citep{2018ApJ...865...46L,2020arXiv200201926Y}, the magnetic field strength \citep{2020arXiv200207996L} and the sonic Mach number \citep{2018arXiv180200024Y}, as well as identifying shocks \citep{IGs} and self-gravitating regions in molecular clouds \citep{Hu20}. In this work, we apply GT to the hot gas in galaxy clusters. For predictions of the POS magnetic field morphology, we focus on the brightest galaxy clusters that have been deeply observed with the {\it Chandra} observatory, namely the Perseus, Virgo/M87, and Coma clusters. We also include the analysis of the synchrotron emission of Perseus. We predict the spatial distribution of the magnetic fields in the clusters, showing that they predominately follow the large structures in the ICM. In particular, the magnetic field follows Perseus' sloshing arms, which agrees with predictions the numerical simulations \citep{2011ApJ...743...16Z}. Although the application of the GT technique on clusters provides promising results, future direct measurements are indispensable for robust conclusions. 
	
	In what follows, we describe the observational data used in this work in detail in \S~\ref{sec:data}. In \S~\ref{sec:theory}, we illustrate the theoretical foundation of the GT in MHD turbulence. In \S~\ref{sec:method}, we describe the algorithm in calculating the magnetic fields using GT. In \S~\ref{sec:result}, we show our predictions of the magnetic field morphology in Coma, Perseus, and M87. In \S~\ref{sec:diss}, we discuss the uncertainties and prospect of GT. We conclude this work in \S~\ref{sec:conc}.

	\section{Observational data}
	\label{sec:data}
	\subsection{X-ray Observation}
	For all clusters in our analysis, we use deep {\it Chandra} observations that are available in the archive. We process the data following standard algorithms \citep{2005ApJ...628..655V} and produce a mosaic image of each cluster, corrected for exposure and vignetting effects, and subtracting the background. After removing the point sources, we find the best-fitting spherically-symmetric  $\beta-$model that describes the X-ray surface brightness distribution. Dividing the initial images by this model, we obtain residual images of gas perturbations in each cluster. The images are produced in the soft X-ray band, i.e., we remove photons with the energies above $3-4$ keV. The center of all cool-core clusters (Perseus, M 87, A2397) coincides with the central AGN. For the only non-cool-core cluster in our sample, the Coma cluster, we choose (R.A., Dec.) = (12h59m42.67, +27$^\circ$56’40.9) (J2000) as the center. For the Coma offset region, we shift the center by $\sim 120$ kpc to account for the large-scale asymmetry at large distances from the center. The details of data preparation are discussed in \citet{2014Natur.515...85Z} and  \citet{IZ19}.
	
	\begin{figure*}[t]
		\centering
		\includegraphics[width=0.78\linewidth,height=0.5\linewidth]{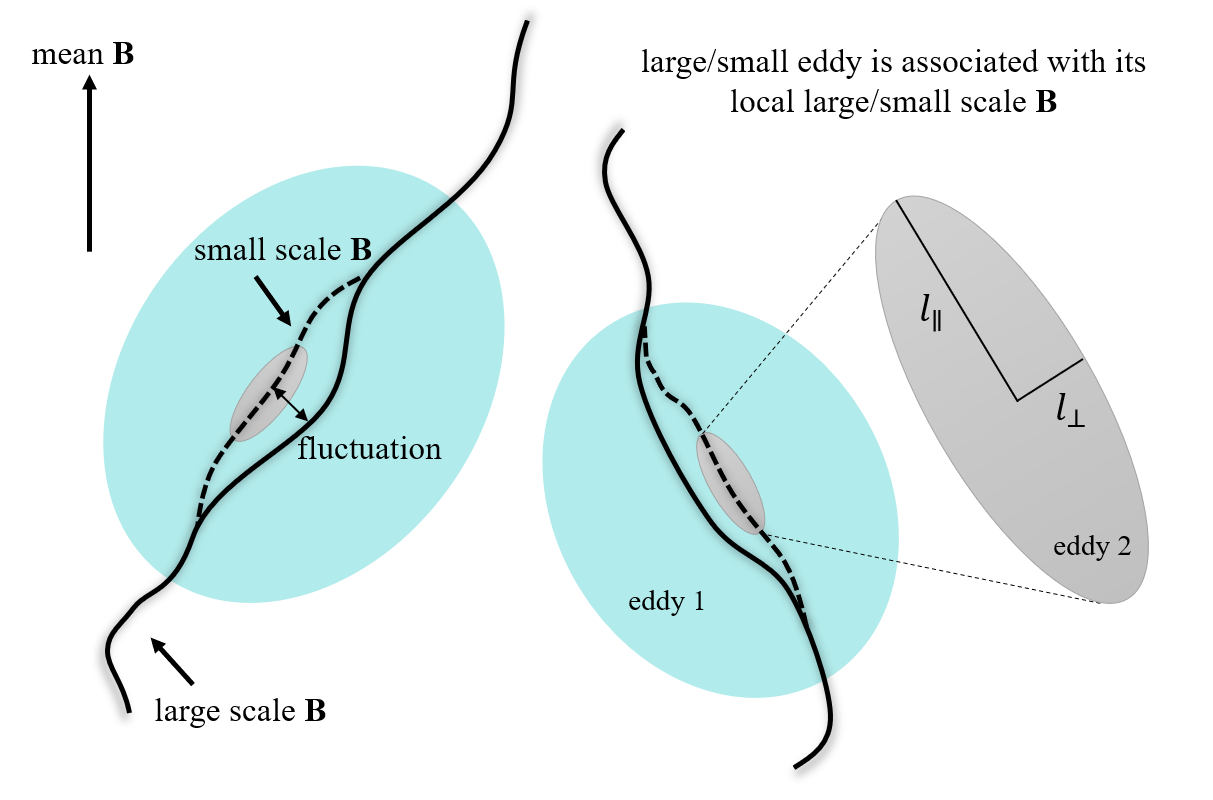}
		\caption{\label{fig:VGT} For trans-Alfv\'{e}nic turbulence, large eddy 1 is less anisotropic since they have similar semi-major axes ($l_\parallel$) and  semi-minor axes ($l_\bot$). Smaller eddy 2 has a relatively larger semi-major axis to the semi-minor axis ratio. Therefore, they are relatively more elongated. The solid curve defines the directions of the local mean magnetic field line \textbf{B} for eddy 1, while the dashed curves define the directions of the local mean magnetic field line for eddy 2 and other small scale eddies. While large eddies induce the global change of the magnetic field, the small eddies still follow the local magnetic field. Based on this property, the density/velocity gradients of the eddy are tracing the projected local magnetic field rather than the global mean magnetic field. For super-Alfv\'{e}nic turbulence, the large scale eddy is isotropic, while the small scale motions are anisotropic and they dominate the contribution to the observed gradients.}.
	\end{figure*}
	\subsection{Radio Observation}
	\label{subsec:SPG}
	Galaxy clusters also harbour large diffuse synchrotron emission sources that are not directly associated with cluster radio galaxies \citep{2019SSRv..215...16V}. In this work, we use the Perseus cluster radio data observed by the Karl G. Jansky Very Large Array (JVLA) in the P-band (230–470 MHz). The observations used were performed for five hours in the B configuration (see \citealt{2017MNRAS.469.3872G} for details). The resulting image has a synthesized beamwidth of $22.1''\times11.3''$ with a grid resolution $3''\times3''$ per pixel and a RMS noise of 0.35 mJy beam$^{-1}$. In addition to the noise level, there is a systematic error due to possible flux scale errors of $\sim 10\%$. These observations have revealed a wealth of inner structures in the mini-halo, a type of faint and diffuse radio structure with steep spectra, filling the cooling core of some relaxed clusters (e.g. \citealt{2019ApJ...880...70G}). In our analysis, we mask low-intensity pixels below the $5\sigma = 1.75$ mJy/beam level.
	
	\section{Theoretical consideration}
	\label{sec:theory}
	\subsection{MHD turbulence theory}
	\label{subsec:turb}
	The GT is developed from the modern MHD theories \citet{GS95} (henceforth GS95) and \citet{LV99} (henceforth LV99). The theory of MHD turbulence has been developing for decades with the understanding of it being anisotropic achieved through theoretical and numerical work \citep{1981PhFl...24..825M,1983PhRvL..51.1484M,1983JPlPh..29..525S,1984ApJ...285..109H}. Introducing the critical balance\footnote{The GS95 study acknowledges that the ”critical balance” between parallel and perpendicular timescales is the key assumption in the derivation of the \citet{1976PhFl...19..134S} equations that was claimed in \citet{Montgomery1982} to describe the anisotropic state of incompressible MHD turbulence. Nevertheless, unlike GS95, the aforementioned papers did not did not provide of the spectra and the scale-dependent ratio between the parallel and perpendicular scales.} between parallel and perpendicular Alfv\'{e}nic motions, GS95 brought the theory of MHD theory to a new stage, in particular obtained the relation between the parallel and perpendicular scales. However, the GS95 anisotropy scaling is derived in the global magnetic field reference frame, in which the predicted scaling is not observable. Later, the study of turbulent reconnection in LV99 demonstrated that the turbulent motions are not constrained by fluid motion perpendicular to the local magnetic field. Indeed, LV99 showed that the time scale for the reconnection for the corresponding eddies equal to the eddy turnover time. As a result, eddies can freely mix  magnetic field parallel to their rotation axes and most of the turbulent energy of the turbulent cascade get cascaded along this path of minimal resistance (see more discussion in  \citealt{2020PhPl...27a2305L}).
	
	In terms of the eddy representation of turbulence that we discussed above, the critical balance is a natural consequence of the freely rotating magnetic eddies inducing Alfv\'{e}n waves parallel to magnetic field, i.e. the eddy period $l_\bot/v_l$ being equal to the period of the generated Alfv\'{e}n wave $l_{\|}/V_A$. Note that $l_\bot$ and $\l_\|$ denote scales of the eddies perpendicular and parallel to the magnetic field, respectively, while $v_l$ is the turbulent velocity of the eddy at the scale $l_{\bot}$ and $V_A$ is the Alfv\'{e}n velocity. Combining the critical balance condition with the constant energy flux condition for the incompressible motions $v_l^2/t_{cas}$ (where the cascading time $t_{cas}\sim l_{\bot}/v_l$), one can get the change of the anisotropy of the eddies with the scale:
	\begin{equation}
	l_\parallel\propto (l_\bot)^{\frac{2}{3}}.
	\label{crit_bal}
	\end{equation}
	The pictorial illustration is given in Fig.~\ref{fig:VGT}. The difference of this scaling from those initially presented in GS95 is that the scales are measured in the system of reference of the eddies rather than with respect to the mean magnetic field. This difference is essential for the Gradient Technique (GT) that we discuss in this paper. The vital importance of using the local magnetic field reference frame in order to observe the universal scale-dependent anisotropy of Alfv\'{e}nic turbulence was demonstrated in \citet{2000ApJ...539..273C, 2002ApJ...564..291C,2001ApJ...554.1175M}.
	
	The turbulence is more complicated in terms of the stochastic density field. For super-sonic turbulence, i.e. for turbulence with $V_L>c_s$, where $c_s$ is the sound velocity, shocks create clumps that make the spectrum of density shallow \citep{2005ApJ...624L..93B,2007ApJ...658..423K}. However, as it was first noticed in \citet{2005ApJ...624L..93B} the low amplitude density fluctuations are elongated and follow the critical balance given by Eq. (\ref{crit_bal}). For galaxy clusters, the turbulence is usually sub-sonic, i.e. $M_s<1$ and therefore one expects the density to follow the velocity scaling. Note that in galaxy cluster outskirts, the turbulence can be supersonic and compressive effects could be non-negligible (see the discussion in \S~\ref{sec:diss}).
	
	\subsection{Transitional scale of turbulence cascade}
	The discussion of MHD turbulence above deals with trans-Alv\'{e}nic turbulence, i.e. with the turbulence that is injected at the injection scale $L$ with the Alfv\'{e}n velocity $V_A$. The case of sub-Alfv\'{e}nic turbulence, i.e. when the Alfv\'{e}n Mach number $M_A=V_L/M_A<1$, where the injection velocity $V_L$ is less than $V_A$, is more complicated, as initially eddy motions are not possible, and the turbulence gets into the critical balance only starting from a smaller scale $l_{trans}\approx L_{inj} M_A^2$, while at the scales $[l_{trans}, L_{inj}]$ the turbulence is {\it weak} \citep{LV99,2000JPlPh..63..447G}. Nevertheless, the sub-Alfv\'{e}nic turbulence is always anisotropic.
	
	For galaxy clusters we deal with super-Alfv\'{e}nic turbulence, i.e. with $M_A>1$ turbulence. For such turbulence the motions at the injection scale are hydrodynamic due to the relatively weak back-reaction of the magnetic field. However, as the kinetic energy of turbulent motions follows the nearly isotropic Kolmogorov cascade, i.e. $v_l^2\sim l^{2/3}$, the importance of magnetic back reaction gets more and more stronger.
	Eventually, at the scale $l_A\approx L_{inj}M_A^{-3}$, the turbulent velocity gets equal to the Alfv\'{e}n velocity \citep{2006ApJ...645L..25L}. For scales less than $l_A$ the turbulence has the MHD nature showing the GS95 anisotropic scaling relation and the corresponding scaling of velocity peturbation:
	\begin{equation}
	l_\parallel\simeq l_A(\frac{l_\bot}{l_A})^\frac{2}{3}
	\end{equation} 
	\begin{equation}
	\label{eq:KS}
	v_{l}\simeq V_L(\frac{l_{\perp}}{L_{inj}})^{\frac{1}{3}}M_A^{\frac{1}{3}}.
	\end{equation}
	Once the telescope resolves the cluster scale smaller than $l_A$, we can expect the density gradients are perpendicular to their local magnetic fields. To get rid of the contribution from large scale eddies, one possible solution is the k-space filter. After the transformation of spatial intensity map to k-space, large scale components, i.e., $l>l_A$, correspond to small k number. \citet{YL17b} numerical showed that by filtering out the components with small k value, there is only contribution from small scale structures left. 
	
	In reality, instead of being isotropic, turbulence on large scales\footnote{On scales larger than the Ozmidov scale \citep{1992JMS.....3..417O}.}  can be anisotropic in galaxy clusters due to the presence of stratification in cluster atmospheres \citep{2014ApJ...788L..13Z,2020MNRAS.493.5838M}. Considering the flux freezing condition, the weak magnetic field, is expected to follow the radial direction of gravity simultaneously. Coincidentally, the density gradient in this case is still perpendicular to the magnetic field showing a different scaling relation from Eq.~\ref{eq.4}. Equivalently, this anisotropy due to gravity on large scale can be considered as an amplification of $l_A$.

	The resolution of X-ray and synchrotron data shall resolve the scales smaller than $l_A$ to implement GT. To figure out the transitional scale $l_A$ for galaxy clusters, we use subsonic turbulence with $M_s\simeq 0.2$ and velocities of the order of $v \simeq 100 - 200$ km/s within the cool cores \citep[e.g.,][]{2007MNRAS.378..245B,2018PASJ...70...11H,2018ApJ...865...53Z}.
	Using the reference parameters $(c_s/v_A)^2\simeq100$, the topical Alfv\'{e}n Mach number of a cool-cluster is then $M_A\simeq10\cdot M_s\simeq2$ \citep{2007MNRAS.378..245B}. As for the global systems, the injection scales with sloshing L$_{inj}$ varies from 20kpc to 200kpc, depending on whether the injection of energy by galaxy merge or galaxy wakes is considered \citep{Vikhrenko (2011)}. The typical scale $l_A$ for galaxy cluster is therefore in the range of 2.5 kpc $\sim$ 25 kpc. This value varies for different objects, for instance, in the Coma central region, the injection scale is in the range of 300 - 400 kpc \citep{2012MNRAS.421.1123C}.
	Also, the typical value of $l_A$ should be larger when taking gravity into account. For our observational data, the smallest scale that can be resolved in Perseus cluster is $\approx$ 6 - 10 kpc, while $\approx$ 4 - 8 kpc for the synchrotron emission \citep{2014Natur.515...85Z,2017MNRAS.469.3872G}. As for the Coma and M 87 cluster, we have minimum resolved scale of $\approx$ 10 - 15 kpc and $\approx$ 1 kpc, respectively. Note, the estimates of the minimal scales from the X-ray images account for the combined point spread function at different line of sights within the regions of our interest.  The X-ray and synchrotron data, therefore, is meeting the requirement for using GT. 
	
	\subsection{Scaling of velocity and density gradient}
	
	The anisotropies of turbulence that we discussed above provide a way to find the direction of magnetic field with observations.
	The first demonstration of using the MHD turbulence anisotropy to trace magnetic field with spectroscopic data  performed in \citet{2002ASPC..276..182L}) and it was elaborated further in \citep{2005ApJ...631..320E, 2015ApJ...814...77E}. In this paper we use gradients that have several practical advantages tracing magnetic fields compared to structure functions employed in the aforementioned studies \citep{2018ApJ...865...54Y}. The theoretical relation between the underlying anisotropic turbulence and the observable parameters is presented in \citep{2012ApJ...747....5L,KLP17b}.      
	
	The GT is based on the properties of MHD turbulence that we discussed above. In particular, gradients induced by both velocity and magnetic fluctuations increase respectively as 
	$v_l/l_\bot\sim l_\bot^{-2/3}$ and $B_l/l_\bot\sim l_\bot^{-2/3}$. The velocity $v_l$ and magnetic field $B_l$ are defined in respect to the {\it local} magnetic field. As a result, the gradients, similar to dust polarization, represent the magnetic field direction averaged along the line of sight. This considerations are at the basis of the GT \citep{2017ApJ...835...41G, YL17a, LY18a, PCA}, SIGs \citep{2018ApJ...855...72L}, and SPGs \citep{2018ApJ...865...59L}.

	As we discussed in \S \ref{subsec:turb} for subsonic turbulence in clusters of galaxies the scaling of density are similar to that of velocity and therefore $\rho_l/l_\bot \sim l_\bot^{-2/3}$. This means that the smallest eddies resolved in observations provide the most important contribution for the intensity gradient signal. 
	Explicitly, since the anisotropic relation indicates $l_\bot \ll l_\parallel$, the velocity gradient scales as \citep{2018arXiv180200024Y}:
	\begin{equation}
	\label{eq.3}
	\nabla v_l\propto\frac{v_{l}}{l_\bot}\simeq \frac{V_L}{L_{inj}}(\frac{l_{\perp}}{L_{inj}})^{-\frac{2}{3}}M_A^{\frac{1}{3}}
	\end{equation}
	here $M_A$ is the Alfv\'{e}n mach number. The direction of velocity gradient is perpendicular to the local directions of magnetic field. Similarly the density gradient can be expressed as:
	\begin{equation}
	\label{eq.4}
	\nabla \rho_l\propto\nabla[\frac{\rho_0v_l}{c_s}\mathscr{F}^{-1}(|\hat{k}
	\cdot\hat{\zeta}|)]=\frac{\rho_0}{c_s}\mathscr{F}^{-1}(|\hat{k}
	\cdot\hat{\zeta}|)\nabla v_l
	\end{equation}
	where $\rho_0$ is the mean density, $\hat{\zeta}$ is the unit vector for the Alfv\'{e}nic mode, fast mode, or slow mode, and $\mathscr{F}^{-1}$ denotes the inverse Fourier transformation. The density $\rho_l$ of the eddy at scale $k$ is expressed as the inverse Fourier transformation of $\rho_k$ \citep{2003MNRAS.345..325C}:
	\begin{equation}
	\rho_l=\mathscr{F}^{-1}( |\rho_k|)=\mathscr{F}^{-1}(\frac{\rho_0v_k}{c_s}|\hat{k}\cdot\hat{\zeta}|)
	\end{equation}
	The scale of density gradient is therefore proportional to velocity gradient with an extra constant term. Eq.~\ref{eq.4} can be further reformed as \citep{2018arXiv180200024Y}:
	\begin{equation}
	\label{eq.6}
	\nabla (\frac{\rho_l}{\rho_0})\propto\frac{\mathscr{F}^{-1}(|\hat{k}\cdot\hat{\zeta}|)}{c_s}\nabla v_l
	\end{equation}
	By assuming the three MHD modes have equivalent velocity, \citet{2003MNRAS.345..325C} deduced the term $\frac{\mathscr{F}^{-1}(|\hat{k}\cdot\hat{\zeta}|)}{c_s}$ can be expressed as:
	\begin{equation}
	\nabla (\frac{\rho_l}{\rho_0})\propto
	\begin{cases}
	M_s \nabla v_l & \text{($\beta\ll1$, \text{slow modes})}\\
	M_A \nabla v_l & \text{($\beta\ll1$, \text{fast modes})}\\
	M_s^2/M_A\nabla v_l & \text{($\beta\gg1$, \text{slow modes})}\\
	M_s^2\nabla v_l & \text{($\beta\gg1$, \text{fast modes})}\\
	\end{cases}
	\end{equation}
	where $M_s$ is the sonic Mach number and $\beta$ is the compressibility.  Similarly, the gradient of $\rho_l^2$ and $v_l^2$ can be expressed as
	\begin{equation}
	\label{eq.grad}
	\begin{aligned}       
	\nabla v_l^2&\propto\frac{v_{l}^2}{l_\bot}\simeq \frac{V_L^2}{L_{inj}}(\frac{l_{\perp}}{L_{inj}})^{-\frac{1}{3}}M_A^{\frac{2}{3}}\\
	\nabla \rho_l^2&\propto\frac{\rho_{l}}{l_\bot}\simeq[\frac{\rho_0}{c}\mathscr{F}^{-1}(|\hat{k}\cdot\hat{\zeta}|)]^2\nabla v_l^2\\
	\end{aligned}
	\end{equation}
	The direction of the gradient is also perpendicular to the local magnetic field.
	
	\begin{figure*}[t]
		\centering
		\includegraphics[width=0.99\linewidth,height=0.64\linewidth]{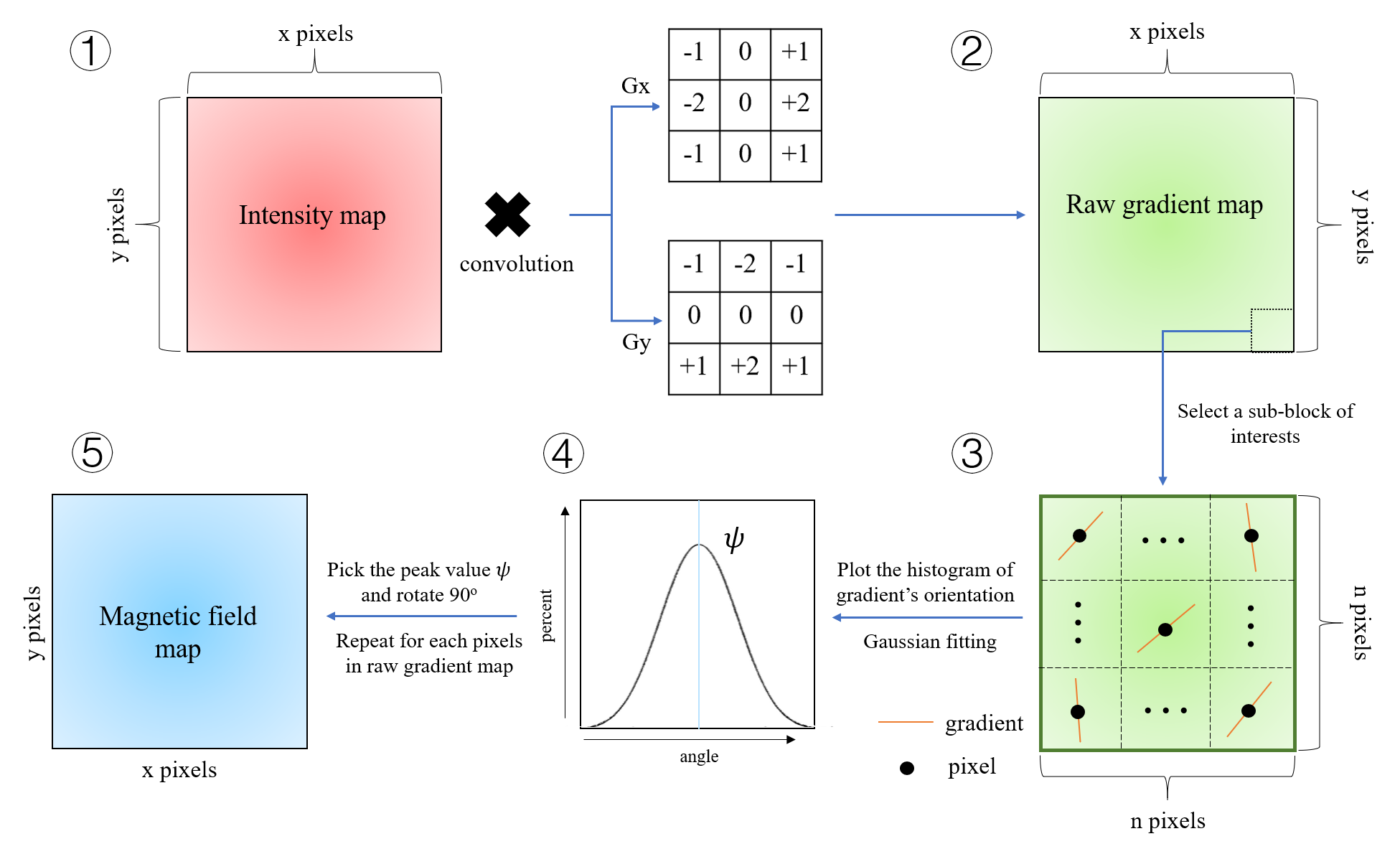}
		\caption{\label{fig:illustration} Diagram of the GT procedure to predict the magnetic field morphology (see \S~\ref{sec:method}). Step 1 is constructing the pixelized 2D gradients map through the convolution of the intensity map (dimension is $x\times y$) with Sobel kernels $G_x$ and $G_y$ (see Eq.~\ref{eq:conv}). In step 2 and step 3, we take a $n\times n$ sub-block from the raw gradient map and plot the histogram of gradient's orientation for the sub-block. A fitting of Gaussian distribution is then applied to the histogram. In step 4, one takes the angle $\psi$ which corresponds to the peak value of the Gaussian to statistically define the mean magnetic field orientation in the corresponding sub-block. Step 2, 3 and 4 are repeated for each pixel in the raw gradient map. As for step 5, we rotate the processed gradient map by 90$^\circ$ getting the projected magnetic field morphology.}
	\end{figure*}
	
	\subsection{Tracing the projected magnetic fields with the gradients}
	
	The mathematical foundations of exploring MHD turbulence and determining the direction of the magnetic field using observations are laid in \citet{2012ApJ...747....5L}. The implementation of this approach for the velocity structure functions is presented in \citet{2019arXiv191002226L}. In Appendix \ref{structure} we summarize the basics of this approach.
	
	The difference between the intensity statistics and velocity/magnetic field statistics is that only the later provide the direct dynamical representation of the physics of turbulence. The properties of densities trace turbulence indirectly and therefore the decomposition into basic MHD modes that is feasible with the spectral lines as well as synchrotron intensities/polarization is not directly applicable for X-ray intensities, that we mostly use in the paper. Nevertheless, for the subsonic turbulence it is possible to treat the density as a passive scalar that reflects basic properties of the velocity field. This justifies our appealing in this paper to the statistical description of gradients presented in Appendix \ref{structure}.

	For an X-ray image of a galaxy cluster in the soft band,we have $I(x,y)\propto\int\rho^2(\Vec{r}) dz$ with $\Vec{r}=(x,y,z)$. Applying the gradient operator $\nabla=(\frac{\partial}{\partial x},\frac{\partial}{\partial y},0)^T$ to $I(x,y)$, the gradient's direction and amplitude can be expressed as:
	\begin{equation}
	\begin{aligned}
	\nabla I(x,y)&\propto\int \nabla\rho^2(\Vec{r}) dz\propto\textbf{U}\cdot\sum_{i=1}^N\nabla_{3D}\rho_i^2(\Vec{r})\\
	|\nabla I(x,y)|&\propto|\int \nabla\rho^2(\Vec{r}) dz|\propto\langle|\nabla_{3D}\rho^2(\Vec{r})|\cos\gamma\rangle\sqrt{N L_{inj}^2}
	\end{aligned}
	\end{equation}
	in which we assume there are $N = L/L_{inj}$ eddies along LOS with distance $L$. The 3D gradient operator is $\nabla_{3D}=(\frac{\partial}{\partial x},\frac{\partial}{\partial y}, \frac{\partial}{\partial z})^T$ and $\gamma$ is the relative angle between the gradient and the POS with $\cos\gamma=|\nabla\rho^2(\Vec{r})|/|\nabla_{3D}\rho(\Vec{r})|$. Note this summation of gradient amplitude as random walk is only valid when $\tan(\gamma/4)>M_A/\sqrt{3}$ \citep{2020arXiv200207996L}. 
	\textbf{U} is the projection operator:
	$$
	\textbf{U}=\begin{pmatrix} 
	+1 & 0 & 0 \\
	0 & +1 & 0 \\
	0 & 0 & 0
	\end{pmatrix}
	$$
	The direction of $\nabla I(x,y)$ is therefore the vector summary of the density gradient along LOS, which is perpendicular to the projected magnetic field. In this work, we rotate the intensity gradient by 90$^\circ$ to predict the magnetic filed orientation unless specified.

	
	\section{Method}
	\label{sec:method}
	\subsection{Implimentation of Gradients Technique}
	\subsubsection{Calculation of intensity gradient}
	In terms of the X-ray and synchrotron observations, the intensity information is obtained from the X-ray surface brightness $I_x$ \citep{1999MNRAS.307..841C} and synchrotron emission intensity $I_\nu$ \citep{1965ARA&A...3..297G}:
	\begin{equation}
	\label{inten}
	\begin{aligned}
	I_x&\propto\int n_e^2\Lambda_edz\\
	I_\nu&\propto\int B_\bot^{1-\gamma}\nu^{\gamma}dz\\
	\end{aligned}
	\end{equation}
	where $n_e$ is the electron number density, $\Lambda_e$ is the X-ray spectral emissivity of the cluster gas due to thermal Bremsstrahlung emission within a certain energy band $\Delta E$. $\Lambda_e$ is negligible here since it is almost independent of temperature in the soft X-ray images (0.5-3.5 keV). $B_\bot$ corresponds to the magnetic field component perpendicular to LOS, the latter given by the z-axis, $\gamma$ is a spectral index of the electron distribution $N(E)\propto E^{2\gamma-1}$, and $\nu$ is the radiative frequency. The X-ray map is further processed to a residual map, i.e., the initial image divided by the best-fitting spherically symmetric $\beta$-model of the surface brightness then minus one. 
	
	As discussed in Appendix \ref{structure} the gradients of synchrotron and X-ray fluctuations are related to the structure functions of the corresponding intensities. The complication that one faces with Eq. (\ref{inten}) is that the the X-ray intensity is proportional to squared density, and a fractional power of $B$ enters the expression enters the synchrotron intensity. The latter problem was addressed in \cite{2012ApJ...747....5L} where it was shown that the structure functions of the synchrotron intensities with $B^{1-\gamma}$ can be successfully represented as a product of the known function of $\gamma$ and the structure function of integrals having $B^2$ dependence. The scaling properties of the latter as a function of the point separation were defined in \cite{2012ApJ...747....5L} and related to the properties of magnetic field. The effects of the structure functions depending on the squared intensities is a simpler problem and it was discussed in \citet{KLP17b}. These studies provide the theoretical foundations of our further analysis.
	
	The residual map emphasizes the surface brightness fluctuations present in the cluster and the corresponding gradient scales as Eq.~(\ref{eq.6}). We denote both X-ray residual map and synchrotron emission map as \textbf{I(x,y)} in calculating the intensity gradients (IGs). As illustrated in Fig.~\ref{fig:illustration}(1), the pixelized gradient map $\psi_{g}$ are calculated from:
	\begin{equation}
	\label{eq:conv}
	\begin{aligned}
	\bigtriangledown_x I(x,y)=G_x * I(x,y)  \\  
	\bigtriangledown_y I(x,y)=G_y * I(x,y)  \\
	\psi_{g}(x,y)=tan^{-1}[\frac{\bigtriangledown_y I(x,y)}{\bigtriangledown_x I(x,y)}]
	\end{aligned}
	\end{equation}
	where $\bigtriangledown_x I(x,y)$ and $\bigtriangledown_y I(x,y)$ are the x and y components of gradients respectively. $*$ denotes the convolution with 3 $\times$ 3 Sobel kernels $G_x$ and $G_y$:
	$$
	G_x=\begin{pmatrix} 
	-1 & 0 & +1 \\
	-2 & 0 & +2 \\
	-1 & 0 & +1
	\end{pmatrix},
	G_y=\begin{pmatrix} 
	-1 & -2 & -1 \\
	0 & 0 & 0 \\
	+1 & +2 & +1
	\end{pmatrix}
	$$
	
	\begin{figure*}[p]
		\centering
		\includegraphics[width=1.00\linewidth,height=1.1\linewidth]{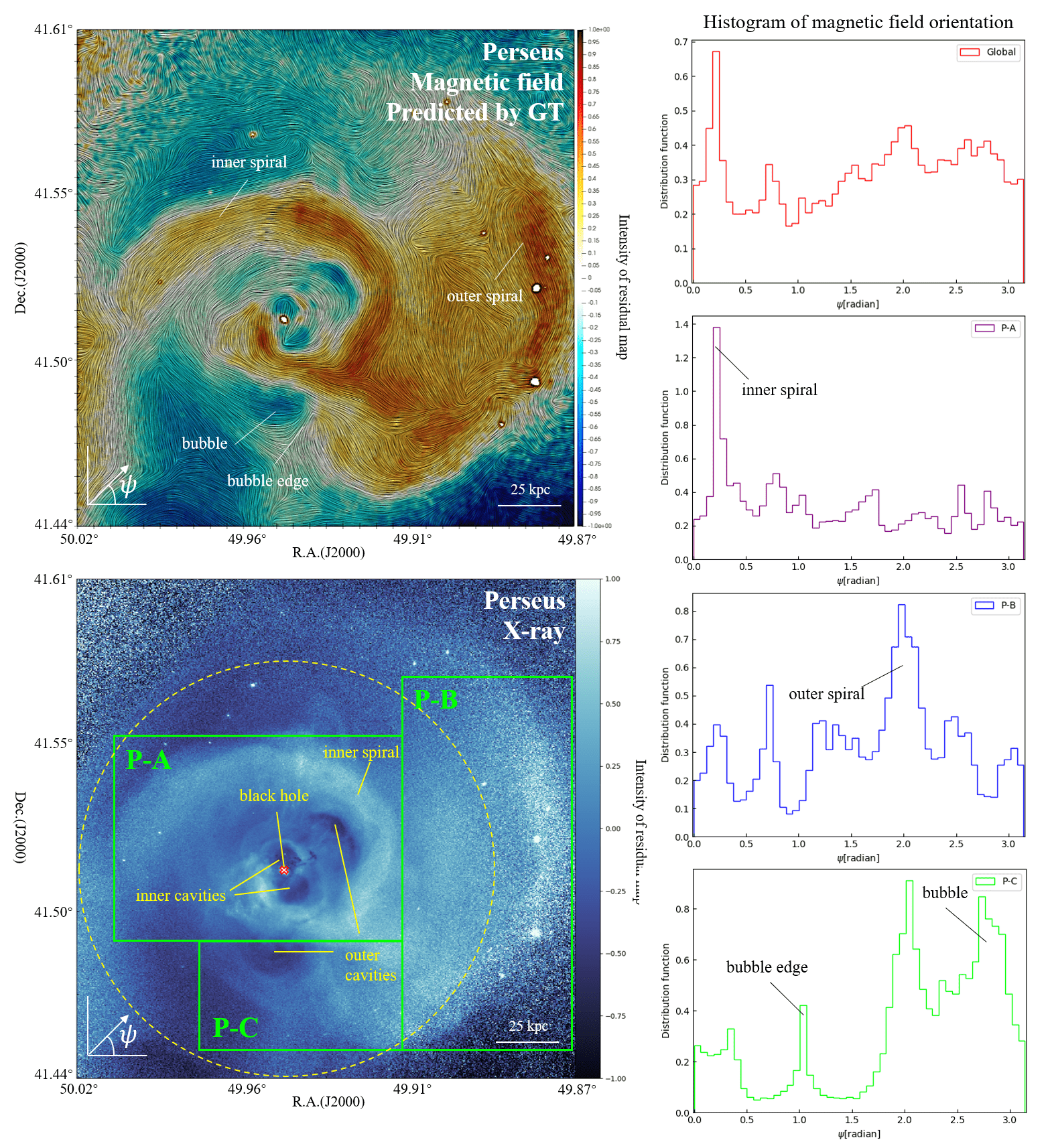}
		\caption{\label{fig:perA}\textbf{Left top:} the predicted magnetic field morphology of the Perseus cluster from GT. The magnetic field is superimposed in the residual map (i.e., the initial image divided by the best-fitting spherically symmetric $\beta$-model of the surface brightness then minus one) using LIC. \textbf{Left bottom:} the residual image of the Perseus cluster. The cluster is divided into three sub-regions, i.e., P-A, P-B, and P-C. \textbf{Right:} the histogram of global magnetic field orientation $\psi$ for the Perseus cluster (top 1st, red) and the histograms of magnetic field orientation $\psi$ for the sub-regions P-A (top 2nd, purple), P-B (top 3rd, blue), and P-C (bottom, green) respectively. The dashed circles corresponds to $4'\approx 80$ kpc. 
		}
	\end{figure*}

	Owing to the fact that the orientation of turbulent eddies with respect to the local magnetic field is a statistical concept, the resulting raw gradient map $\psi_{g}(x,y)$ is not necessarily required to have any relation to the local magnetic field direction \citep{YL17a}. The perpendicular relative orientation only appears when the gradient sampling is enough. Ideally, the distribution of intensity gradient's orientation within a sampled sub-region is Gaussian. The peak of the Gaussian distribution reflects the statistically most probable gradient orientation in the sub-region. By rotating the most probable gradient orientation with 90$^\circ$, one can have the local mean magnetic field direction, see Fig.~\ref{fig:illustration}(2-4). This procedure proposed in \citet{YL17a} is called the sub-block averaging method.
	
	\subsubsection{Adaptive sub-block averaging method}
	\label{subsec:asb}
	The critical steps of the sub-block averaging method are: (i) selecting a sub-region with an appropriate size; (ii) plotting the histogram of gradients' orientation within this sub-region; (iii) taking the angle of orientation corresponding to the Gaussian fitting peak value of the histogram. The resultant angle statistically defines the mean gradient's orientation in the corresponding sub-region. The original recipe used in \citet{YL17a} divides the entire gradient map $\phi_g(x,y)$ into a few amounts of pieces with identical size. For instance, a regular grid 512$\times$512 map can be divide into 16 pieces having 32$\times$32 pixels for each. Different from  \citet{YL17a}, \citet{Hu20} take each pixel of $\psi_{g}(x,y)$ as the center of a rectangular sub-block and applying the recipe of sub-block averaging. The size of each sub-block is not identical but determined by the Gaussian fitting error. When the sampling is insufficient, the fitting algorithm outputs a significant fitting error. We vary the sub-block size from large to small for each pixel and check the corresponding fitting errors. The optimal sub-block size is selected when the error gets its minimum value. Since the new sub-block averaging method automatically determines the sub-block size for each pixel, we denote it as an adaptive sub-block averaging (ASB) method. In this work, we follow the ASB method.
	
	Note although the adaptive sub-block averaging outputs the per-pixel magnetic field map, the effective resolution is determined by the sub-block size in the corresponding position. The minimum sub-block size is empirically setting as 20$\times$20 \citep{YL17a, LY18a} so that the maximum effective resolution of GT is $\approx20''$ for all X-ray data and $\approx1'$ for synchrotron data. We denote the gradient's orientation processed by the adaptive sub-block averaging as $\psi_A(x,y)$. By rotating the prepared gradient map by 90$^\circ$, we obtain the predicted magnetic field morphology, see Fig.~\ref{fig:illustration}(5).
	
	\subsubsection{Abnormal gradients correction}
	\label{subsub:RHT}
	The accuracy of GT in terms of magnetic field tracing is correlated to the Signal-to-Noise Ratio (SNR) in the observational data \citep{LY18a}. The ASB usually is sufficient to highlight statistically crucial components and suppress parts of noise. However, when the Signal-to-Noise Ratio (SNR) is less than 1, the ASB does not precisely reveal the gradient since the histogram of gradient's orientation tends to be a uniform distribution rather than Gaussian. To eliminate the contribution from noise, we then introduce the \textbf{Rotational Image Test}. After the first implementation of ASB using raw X-ray map, we rotate the map \textbf{I(x,y)} by 90$^\circ$ getting \textbf{$I_R(x,y)$}. The recipe of gradient's calculation is repeated for \textbf{$I_R(x,y)$} and results in the gradient map $\psi_R(x,y)$. In the case that the noise is insignificant, the Gaussian histogram shall switch 90$^\circ$ and difference between $\psi_A(x,y)$ and $\psi_R(x,y)$ is also 90$^\circ$. However, when SNR is very small, the histogram is still a uniform distribution, which produces a similar fitting angle as $\psi_A(x,y)$. Therefore, we firstly mask the pixel in which the difference between $\psi_A(x,y)$ and $\psi_R(x,y)$ are less than $\pi/2$. The masked pixel is then interpolated based on the neighboring vectors. 
	
	In addition to the rotational histogram test, we employ the construction of pseudo-Stokes parameters to reduce the contribution from noise. In view of that magnetic field lines are continuous, a noise-induced abnormal gradient vector can discontinues the streamline. \citet{Hu20} proposed to smooth the outlying abnormal gradients by constructing pseudo-Stokes parameters $Q$ and $U$:
	\begin{equation}
	\begin{aligned}
	Q(x,y)=I(x,y)\cos(2\psi_A(x,y))\\
	U(x,y)=I(x,y)\sin(2\psi_A(x,y))
	\end{aligned}
	\end{equation}
	The weighting term $I(x,y)$ helps recover the Gaussian properties of the cosine and sine angular components \citep{IGs}. Therefore, instead of smoothing directly the cosine and sine components, the Gaussian convolution is applied to both $Q$ and $U$. The resulting gradient vector is calculated from $\psi(x,y)=\frac{1}{2}\tan^{-1}(U/Q)$. 
	
	\section{Results}
	\label{sec:result}
	In this section, we present the predicted magnetic field maps for the Perseus (the brightest in X-rays), M87/Virgo (the closest) and Coma (the brightest non-cool-core) clusters. We refer the reader to \citet{2011MNRAS.418.2154F,2007ApJ...665.1057F,2013Sci...341.1365S} for deep X-ray observations of these clusters, their interpretation and discussion on prominent structures. 
	The magnetic fields are visualized by the Line Integral Convolution \citep{LIC} and the convention of $\psi$ is defined in Cartesian coordinates, i.e., west-to-north, with the center being the center of the cluster. Before applying GT, we first smooth the X-ray residual maps \textbf{I(x,y)} with a 3 arcsec Gaussian filter, and excise significant point sources in the calculation.

	\begin{figure*}[t]
		\centering
		\includegraphics[width=.99\linewidth,height=1.05\linewidth]{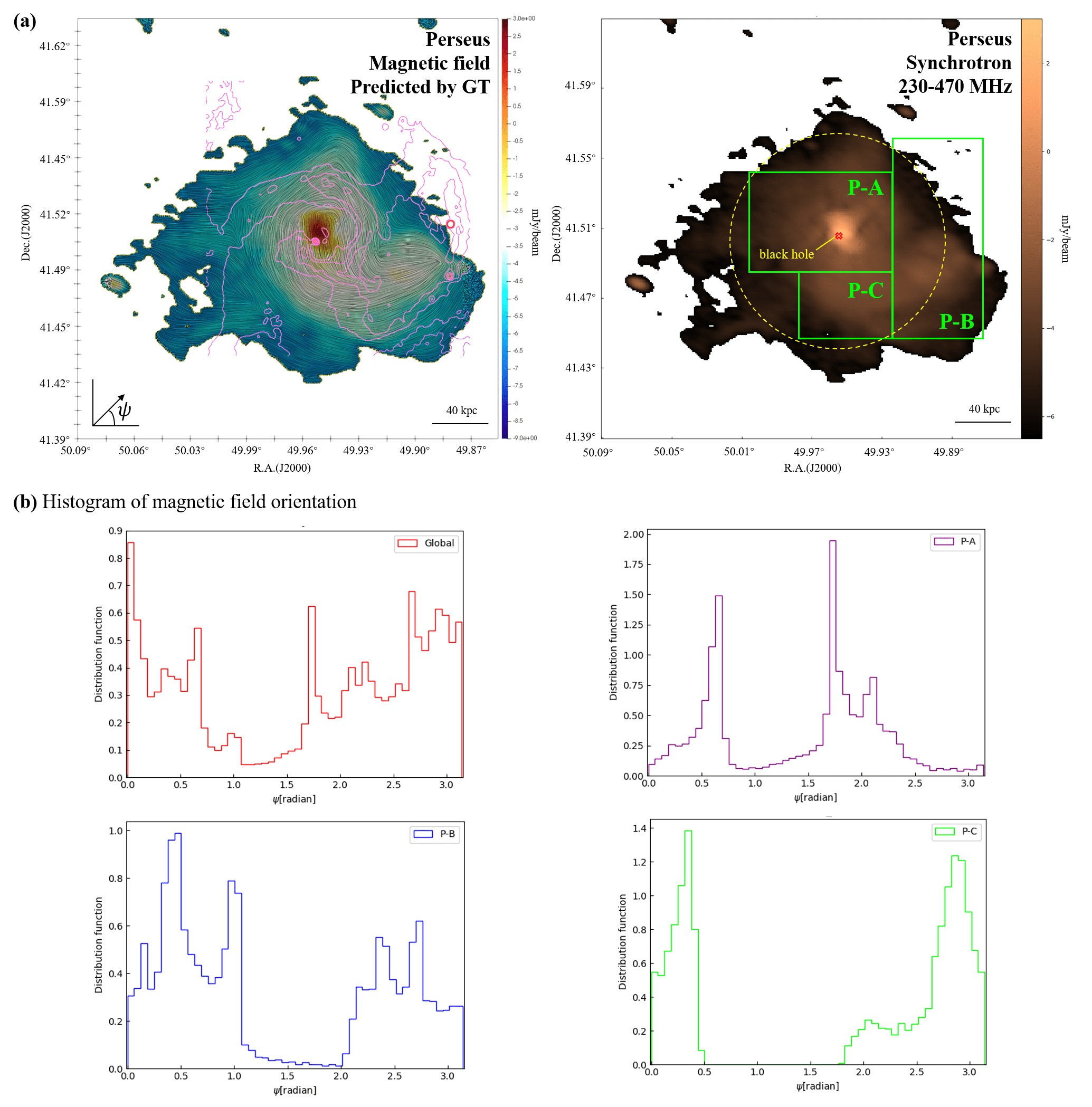}
		\caption{\label{fig:RA}\textbf{Panel a:}  the predicted magnetic field morphology of the Perseus cluster from GT (left) using synchrotron emission data (right). The magnetic field is superimposed on the intensity map with X-ray contours overlaid. Three sub-regions P-A, P-B, and P-C corresponds to the ones highlighted in Fig.~\ref{fig:perA}. \textbf{Panel b:} the histogram of global magnetic field orientation $\psi$ for the global Perseus cluster (top left), sub region P-A (top right), P-B (bottom left), and P-C (bottom right). The dashed circle corresponds to $4'\approx 80$ kpc.
		}
	\end{figure*}
	\subsection{Predicted magnetic field morphology in the Perseus cluster}
	\subsubsection{X-ray observation}
	The predicted POS magnetic field morphology of the Perseus cluster is shown in Fig.~\ref{fig:perA}. We test four Gaussian kernels used in the smoothing of Pseudo-stokes parameters. We select four Gaussian kernel widths $0$, $3''$, $5''$, and $10''$, and then calculate the corresponding angular dispersion of the gradient. We get the angular dispersion $\approx$ 0.991, 0.974, 0.948, and 0.945 in radian unit respectively. In view that the difference between Gaussian kernel widths $5''$ and $10''$ is insignificant and a large kernel may oversmooth the vectors, we choose the width $5''$ as the standard value in the calculation.
	
	One can see in Fig.~\ref{fig:perA} that the magnetic fields are predominantly following the sloshing arm of the cluster. The histogram of global magnetic field orientation shows a significant peak at angle $\approx0.25$ in radian unit, which is also the orientation of the inner sloshing arm. The histogram is drawn in the range of [0, $\pi$) with the bin size 50. Note we do not distinguish between the angle 0 and $\pi$. We outline three sub-regions in the residual map: P-A, P-B, and P-C. P-A contains the inner sloshing arm and the inner bubble structures, while P-B has the outer sloshing arm, and P-C includes an outer bubble. In the histogram of P-A, we find a similar peak at angle $\approx0.25$ as the one in the global histogram, corresponding to the orientation of the inner sloshing arm. This comes from the tangential motion of the plasma (sloshing) that stretched the magnetic field lines \citep[e.g.,][]{2011ApJ...743...16Z}. Note that perturbations in the innermost $\sim 30$ kpc region (within the P-A region), are dominated by the central bubbles and shock-heated gas around them. Therefore, the anisotropy scaling relation could be different from $\nabla \rho_l\propto\frac{\rho_0}{c_s}\mathscr{F}^{-1}(|\hat{k}\cdot\hat{\zeta}|)\nabla v_l$ in this region. Nevertheless, in clusters, we expect the weak magnetic fields to follow the anisotropic direction still. 
	In this case, the resulting direction of the intensity gradient is perpendicular to the magnetic fields, and GT also gives a correct prediction. 
	Additionally, for the brightest part of the inner sloshing (rightmost part of the P-A region) and the outer sloshing (P-B region), we can see similar azimuthal magnetic fields following the sloshing spiral arms. For P-C, the predicted magnetic fields rapidly change directions by 90$^\circ$ at the bubble edge (see the first panel of Fig.~\ref{fig:perA}). 
	As a result, there is a deficit at angle $\approx2.50$ between the two peaks in the histogram, and a small peak at angle $\approx1.00$. 
	The difference between the angle $\approx1.00=57.32^\circ$ and $\approx2.50=143.31^\circ$ is almost 90$^\circ$. This change is consistent with our theoretical expectation of shocks (see \S~\ref{sec:diss}) or magnetic draping caused by the rising bubble \citep{2008ApJ...677..993D}. 
	

	\subsubsection{Radio observations}
	The synchrotron emission from the Perseus cluster is observed by the Karl G. Jansky Very Large Array (JVLA) B-configuration at 230-470 MHz. The data show several structures associated with the mini-halo, which is expected to be influenced both by the AGN activity and the sloshing motion of the hot gas \citep{2017MNRAS.469.3872G}. Fluctuations of synchrotron emission can be used to study properties of the magnetic field, including tracing the magnetic field directions \citep{2012ApJ...747....5L}\footnote{Apart from magnetic field tracing, the mathematical framework in \citet{2012ApJ...747....5L} demonstrates how to separate contributions from fundamental MHD modes (see \citet{2020arXiv200403514C}) and together with \citet{2000ApJ...537..720L} was the basis for the further development in the field of observational studies of anisotropic MHD turbulence from observations (see \citet{2015ApJ...814...77E,KLP17b}.)}. This study presents the foundations for the SIGs introduced in \citet{2018ApJ...855...72L}. Therefore we can use synchrotron emission here to find magnetic field directions. 
	Nevertheless, different from the X-rays emitted by extremely hot gas through the thermal bremsstrahlung emission \citep{1988xrec.book.....S}, synchrotron emission arises from relativistic electrons spiraling along the magnetic fields lines \citep{1970ranp.book.....P}. Due to the different emission sources and mechanisms, the X-ray and synchrotron emissions likely reveal different structures in the ICM. A detailed comparison of the X-ray and synchrotron observations can be found in \citet{2017MNRAS.469.3872G}. In this work, we compare the magnetic fields calculated using these two data sets.
	
	The predicted magnetic field from GT using synchrotron emission is presented in Fig.~\ref{fig:RA}. Low-intensity pixels below the $5\sigma = 1.75$ mJy/beam level are masked in the analysis. The output magnetic field map covers the majority of the X-ray sub-regions P-A, P-B, and P-C (see Fig.~\ref{fig:perA}). Visually, we can see the magnetic field is significantly bent in the high-intensity center and follows the sloshing arm, which is also seen in  Fig.~\ref{fig:perA} of X-ray data. 
	\begin{figure}[t]
		\centering
		\includegraphics[width=0.99\linewidth,height=1.45\linewidth]{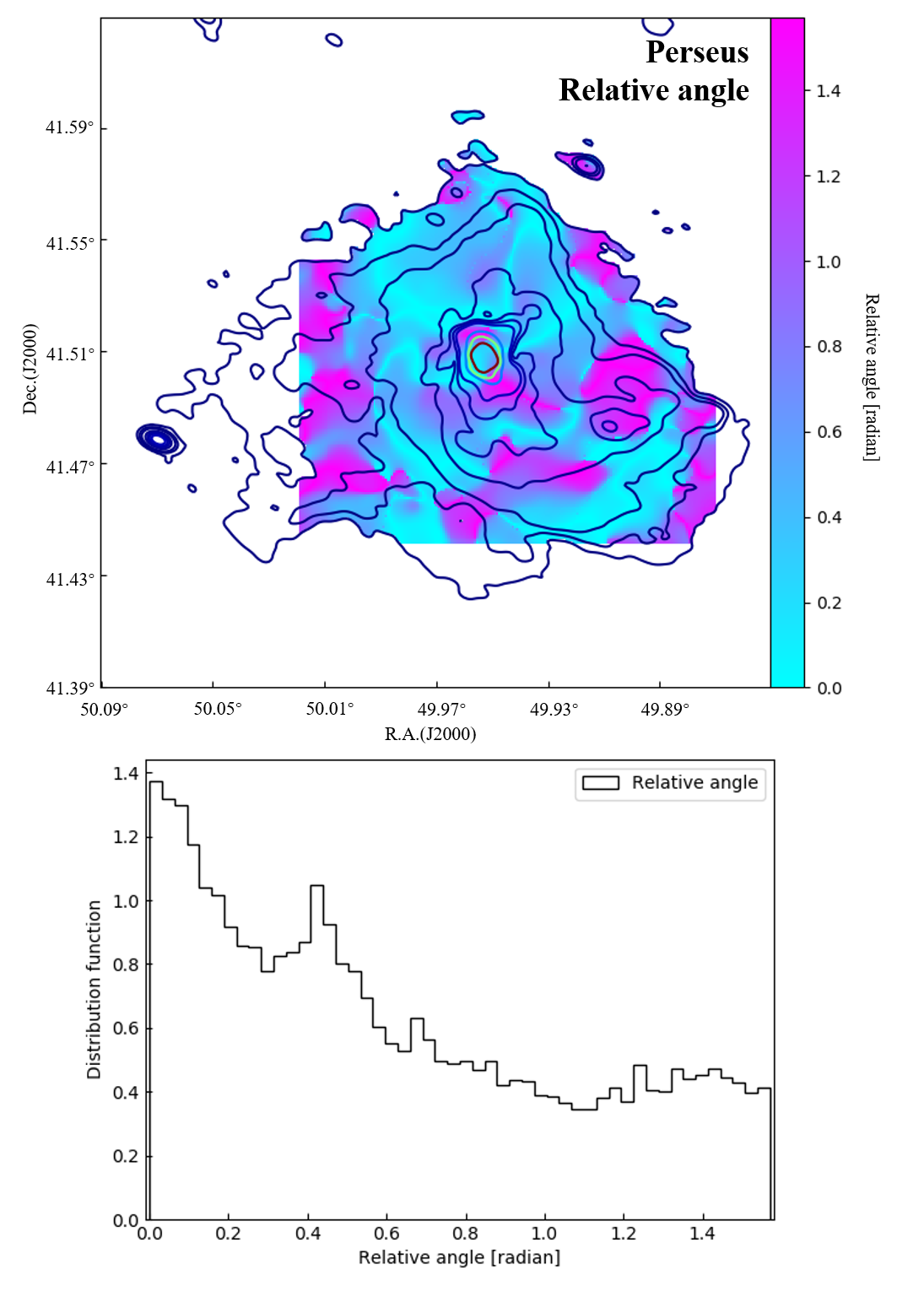}
		\caption{\label{fig:per-GRA} \textbf{Right:} the distribution of the relative angle between the magnetic field inferred from X-ray data and synchrotron emission with synchrotron contours (at 1.75$\times10^{-3}$, 5$\times10^{-3}$, 1$\times10^{-2}$, 3$\times10^{-2}$, 5$\times10^{-2}$, 7$\times10^{-2}$, 0.1, 0.5, 1, 2 mJy/beam levels) overlaid. \textbf{Bottom:} the histogram of the global relative angle between the magnetic fields inferred from X-ray data and synchrotron emission by GT.}
	\end{figure}
	\begin{figure}[t]
		\centering
		\includegraphics[width=0.99\linewidth,height=0.75\linewidth]{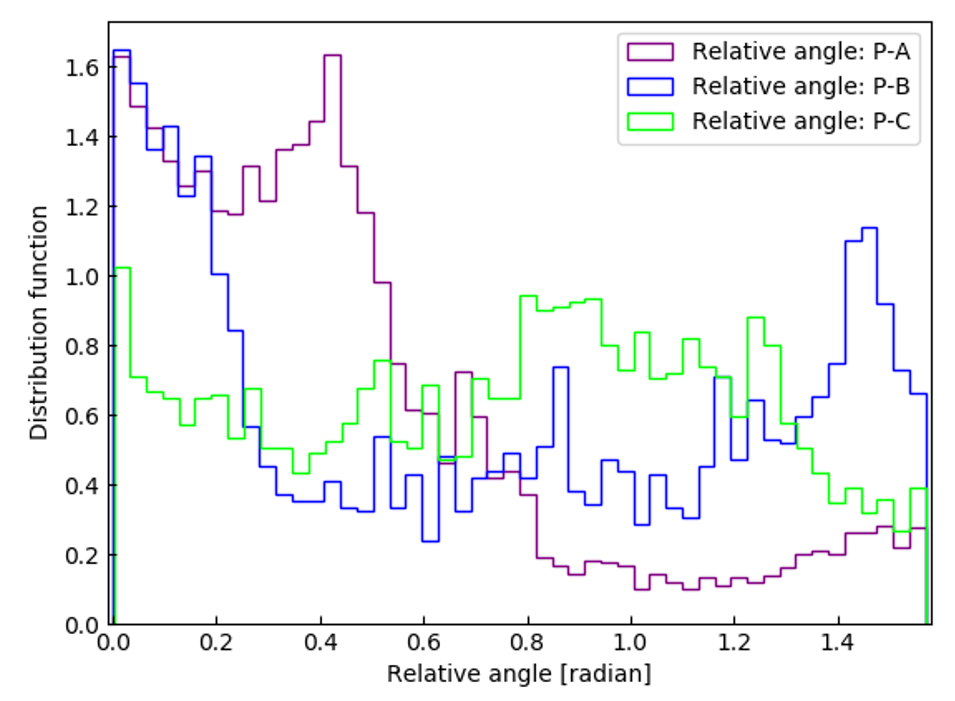}
		\caption{\label{fig:hist_sub}The histogram of the relative angle between the magnetic field calculated from X-ray data and synchrotron data. The histogram is drawn for each sub-region P-A, P-B, and P-C respectively, see Fig.~\ref{fig:perA} and Fig.~\ref{fig:RA} for details of each sub-region.}
	\end{figure}
	We plot the histogram of magnetic field orientation in Fig.~\ref{fig:RA}. For the global magnetic field, the histogram exhibits several peaks at $\psi\approx0$, 0.6 $(\approx34.37^\circ)$, and 1.8 $(\approx103.13^\circ)$, which are very close to the peaks seen in the X-ray data. As for the sub-region P-A, its histogram exhibits two distinct peaks at $\psi\approx$ 0.6 $(\approx34.37^\circ)$ and 1.8 $(\approx103.13^\circ)$. The former corresponds to the sloshing arm, which we have seen in Fig.~\ref{fig:perA}. The latter corresponds to the central region. In terms of the histograms of sub-regions P-B and P-C, we find several similar features that are also seen in Fig.~\ref{fig:perA}, e.g., the two peaks at $\psi\approx$ 0.5 - 1.0 for P-B and $\psi\approx$ 2.9. Although the synchrotron data does not show the peaks at $\psi\approx$ 2.0 in P-B and $\psi\approx$ 1.0 in Fig.~\ref{fig:RA}. Three factors may contribute to the discrepancies between magnetic fields derived from the X-ray and synchroton data: (i) the X-ray and synchrotron emissions are tracing different parts of the ICM; (ii) some structures are not well-resolved due to the relatively lower resolution of the synchrotron data; (iii) the two maps do not fully overlap. For example, the synchrotron data does not cover the upper part of P-B. Also, potentially cosmic ray electrons can introduce external gradients to the results of synchrotron data. This effect of cosmic-ray gradients will be studied elsewhere.
	
	\begin{figure*}[p]
		\centering
		\includegraphics[width=1.00\linewidth,height=1.08\linewidth]{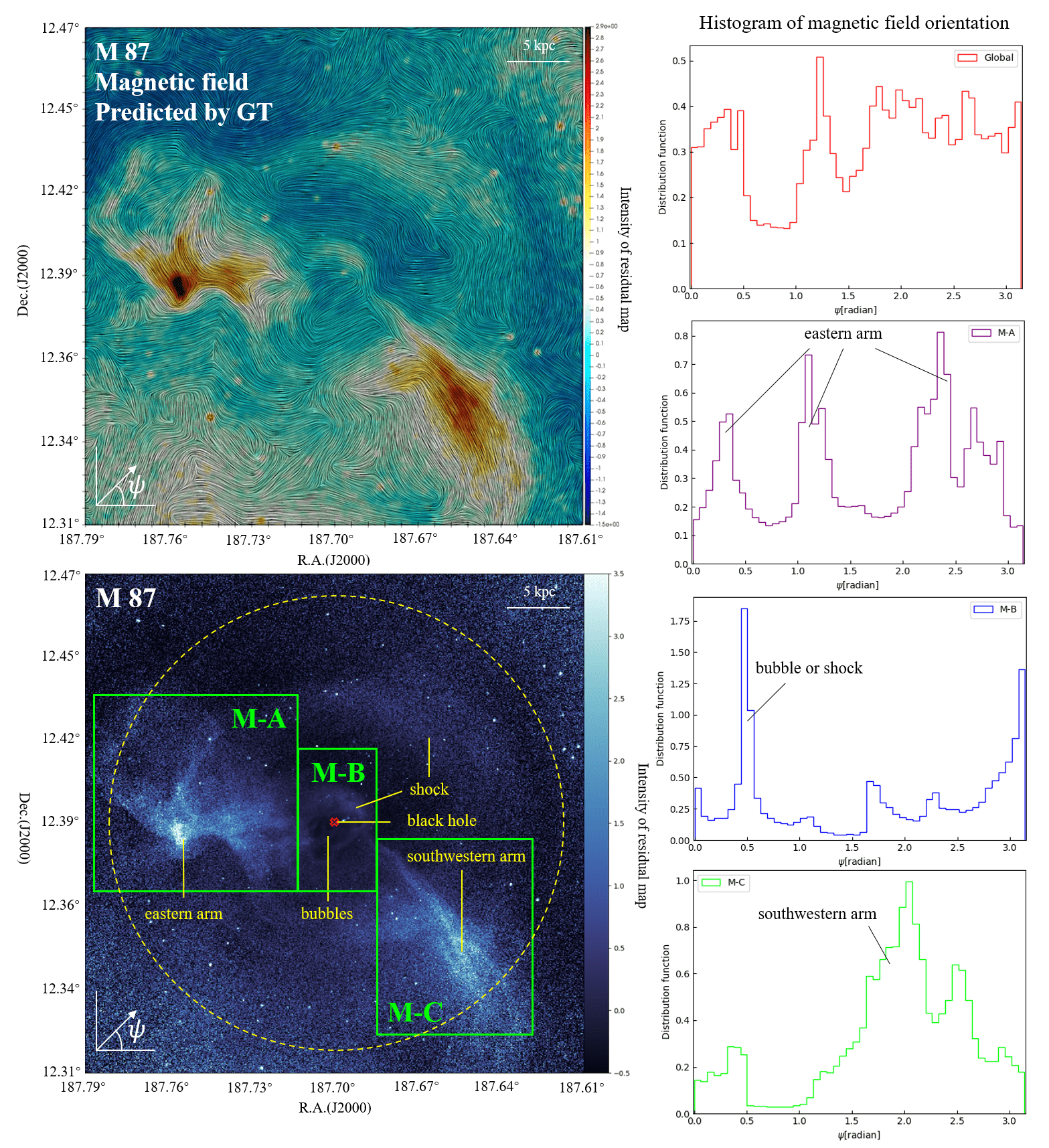}
		\caption{\label{fig:M87}\textbf{Left top:} the predicted magnetic field morphology of the Virgo/M87 cluster from GT. The magnetic field is superimposed in the residual map (i.e., the initial image divided by the best-fitting spherically symmetric $\beta$-model of the surface brightness then minus one) using LIC. \textbf{Left bottom:} the residual image of the cluster. The cluster is divided into three sub-regions, i.e., M-A, M-B, and M-C. \textbf{Right:} the histogram of global magnetic field orientation $\psi$ for the M 87 cluster (top 1st, red) and the histograms of magnetic field orientation $\psi$ for the sub-regions M-A (top 2nd, purple), M-B (top 3rd, blue), and M-C (bottom, green) respectively. The dashed circles corresponds to $4'\approx 20$ kpc. 
		}
	\end{figure*}
	
	We quantitatively compare the magnetic fields derived from the X-ray and the synchronton data first by computing the mean field directions. The details of the calculations can be found in the Appendix~\ref{appendx}. We denote the mean magnetic field calculated from the synchrotron data as $\bar{\mu}_s$, while $\bar{\mu}_x$ is for the X-ray data. We find for sub-regions P-A, P-B, P-C, the value of $\bar{\mu}_s$ is 1.443, 1.260, and 2.065. respectively. The corresponding values of $\bar{\mu}_x$ are 1.434, 1.451, and 2.107. Therefore, in terms of the mean magnetic fields, the X-ray and synchrotron data sets give similar results. 
	
	In addition, we calculate the relative angle between the magnetic fields obtained from the X-ray data and the synchrotron data. The resolution of the synchrotron data is $\approx 3''$ per pixel, while it is $\approx 1''$ per pixel for the x-ray. Therefore, the X-ray data resolves more magnetic field vectors. We first match the two maps in coordinates and take the angular average for every 3$\times$3 magnetic field vectors obtained from the X-ray. The resolution of the magnetic field in the X-ray map is, therefore, reduced to the same as the one in synchrotron data. The resulting relative angle distribution is presented in Fig.~\ref{fig:per-GRA}. We find in the corresponding sloshing arm part and its surroundings, the magnetic field shows general agreement. Since the X-ray and the synchrotron emissions are probably tracing different parts of the ICM, we also see some disagreement. We plot the histogram of the relative orientation in Fig.~\ref{fig:per-GRA}. The histogram is close to a single Gaussian distribution with a peak value at 0 and a standard deviation of $\approx26.93^\circ$. The histograms of the relative angle for P-A, P-B, P-C are plotted in Fig.~\ref{fig:hist_sub}. P-A's histogram shows that the majority of the relative angle is less than 0.5, which indicates a good alignment. P-B gives the relative angle either parallel (relative angle $\approx$ 0) or perpendicular (relative angle $\approx\pi/2$). While for P-C, the alignment is more close to random. 
	
	To further quantify the agreement between the two magnetic field maps, we introduce the \textbf {Alignment Measure} (AM)\footnote{This measure was borrowed from the grain alignment theory (see \citealt{2007JQSRT.106..225L}) and introduced to characterize the velocity gradient alignment in respect to magnetic field in \citet{2017ApJ...835...41G}. Later it was borrowed by other groups (see \citealt{2019A&A...622A.166S,2019ApJ...887..136C}) to characterize the alignment of the features that they study with the magnetic field.}:
	\begin{equation}
	\label{eq.AM}
	AM=2(\langle cos^{2} \theta_{r}\rangle-\frac{1}{2})
	\end{equation}
	where $\theta_r$ is the relative angle of individual pixels, while $\langle ...\rangle$ denotes the average within a region of interest. In the case of a perfect global agreement between the two vector maps, we have AM = 1, while AM=-1 indicates that the global relative angle is 90$^\circ$. The standard error of the mean gives the uncertainty $\sigma_{AM}$, which is negligible with sufficiently large samples.  
	
	For the global magnetic field, we get AM $\approx0.29$, while the AM $\approx0.53$, 0.10, 0.02 for P-A, P-B, and P-C respectively. As a check of chance alignment between predicted magnetic field maps, we perform three Monte Carlo analyses by randomly selecting 10000, 20000, and 30000 vectors from each map (note the magnetic field map from X-ray is reduced to the same resolution as synchrotron). The resulting histograms of relative angle appear as uniform distribution with AM $\approx$ 0.032, 0.028, 0.022, which rejects the hypothesis of chance alignment.
	
	\subsection{Predicted magnetic field morphology in M 87}
	
	We also apply the GT analysis to the X-ray residual maps of M 87, the central galaxy of the M 87 cluster. Similarly to our analysis of Perseus, we outline three sub-regions, i.e., M-A, M-B, and M-C. M-A and M-C both contain eastern and southwestern arms, while M-B includes the inner bubble and jet. Fig.~\ref{fig:M87} shows the predicted magnetic field morphology, X-ray residual map, and the histograms of magnetic field orientation. The global histogram exhibits rather complex features, owing to the complicated structures in M 87. The histogram of M-A shows three distinct peaks at angle $\approx0.25$, $1.20$, and $2.40$. These three peaks correspond to the three features in the eastern arm in M-A. There is only one significant outflow arm in M-C and hence the corresponding histogram gives a single peak at angle $\approx2.00$. As for the M-B, the histogram shows a single peak at $\approx0.50$. Also, our the results show coherent magnetic field along the shock (above the M-B and M-C regions). A similar analysis is also repeated for Abell 2597, see the Appendix~\ref{ap.A} for details and results. 
	
	\begin{figure*}[t]
		\centering
		\includegraphics[width=1.0\linewidth,height=.92\linewidth]{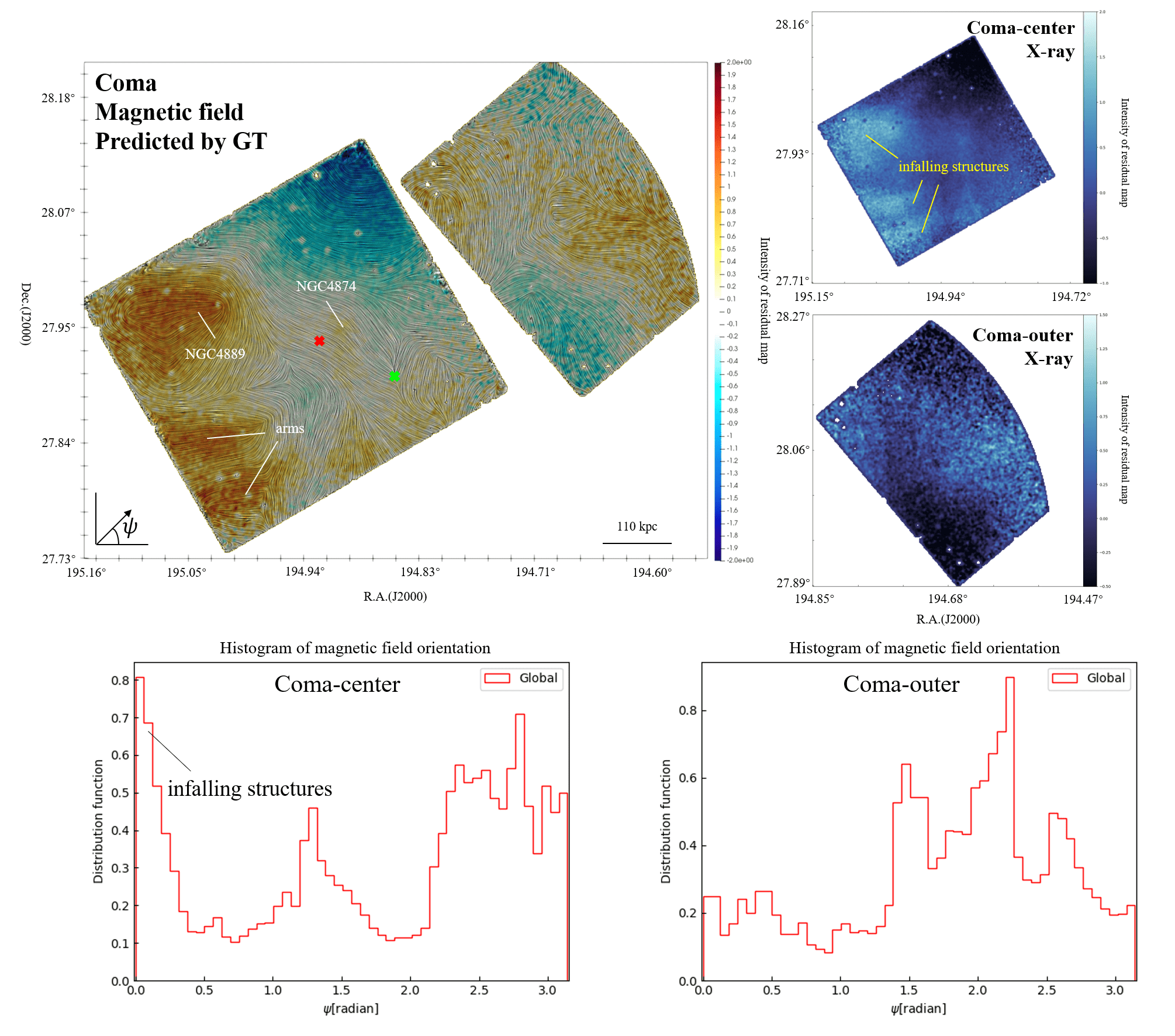}
		\caption{\label{fig:coma}\textbf{Top left:} the predicted magnetic field morphology of the Coma cluster center region and outer region from GT. The magnetic field is superimposed in the residual map (i.e., the initial image divided by the best-fitting spherically symmetric $\beta$-model of the surface brightness then minus one) using the LIC. \textbf{Bottom:} the histograms of global magnetic field orientation $\psi$ for the Coma cluster's center region (left) and outer region (right). The red cross denotes the center R.A.=12h59m42.67 (J2000) and Dec.=+27$^\circ$56$'$40.9$''$ (J2000) used for the $\beta$-model fitting in Coma-center and green cross denotes the center R.A.=12h59m22.67 (J2000) and Dec=+27$^\circ$54$'$40.9$''$ (J2000) used for Coma-outer. 
		}
	\end{figure*}
	
	\subsection{Predicted magnetic field morphology in the Coma cluster}
	The Coma cluster is a well-studied nearby massive cluster that has undergone several recent mergers with intermediate-mass subgroups \citep{1997ApJ...474L...7V}. 
	Unlike the other three clusters with cool cores and active central supermassive black holes, Coma is a merging, non-cool-core cluster without AGN activity in the center. In this work, we use two sub-regions of the Coma cluster which were observed by {\it Chandra}. Following \citet{IZ19}, we denote these two sub-regions as Coma-center and Coma-outer. The X-ray residual maps are shown in Fig.~\ref{fig:coma}. 
	
	The GT-predicted magnetic field of Coma is shown in Fig.~\ref{fig:coma}. 
	In general, the predicted magnetic field is following the structures in the ICM as expected, and in high-intensity regions, the change of the magnetic field orientation is more rapid. In Fig.~\ref{fig:coma}, we plot the histogram of the magnetic field orientation $\psi$.  For the Coma-center region, the histogram appears bimodal, with two peaks at $\psi\approx 1.4$ radian $= 80^\circ$ and $\psi\approx 0^\circ$. Since these two regions in Coma do not have strong shocks which potentially could change the gradient's direction (see \S~\ref{sec:diss}), the bending of the magnetic fields is likely caused by subsonic bulk motion. Indeed, Coma-center has two structures associated with the gas stripped from merging subclusters (seen as orange enhancements in Fig.~\ref{fig:coma},  the largest enhancement is associated with two massive galaxies) \citep{2013Sci...341.1365S}. The magnetic fields appear to be oriented around one of them, likely as a result of magnetic draping.
	
	Unlike Coma-center, the histogram of Coma-outer is approximately a single Gaussian distribution with a peak value $\psi\approx 2.0$ radian $=114.5^\circ$. This suggests a lack of strong perturbations in the magnetic fields in Coma-outer on probed scales.
	

	\begin{figure}[t]
		\centering
		\includegraphics[width=0.99\linewidth,height=2.1\linewidth]{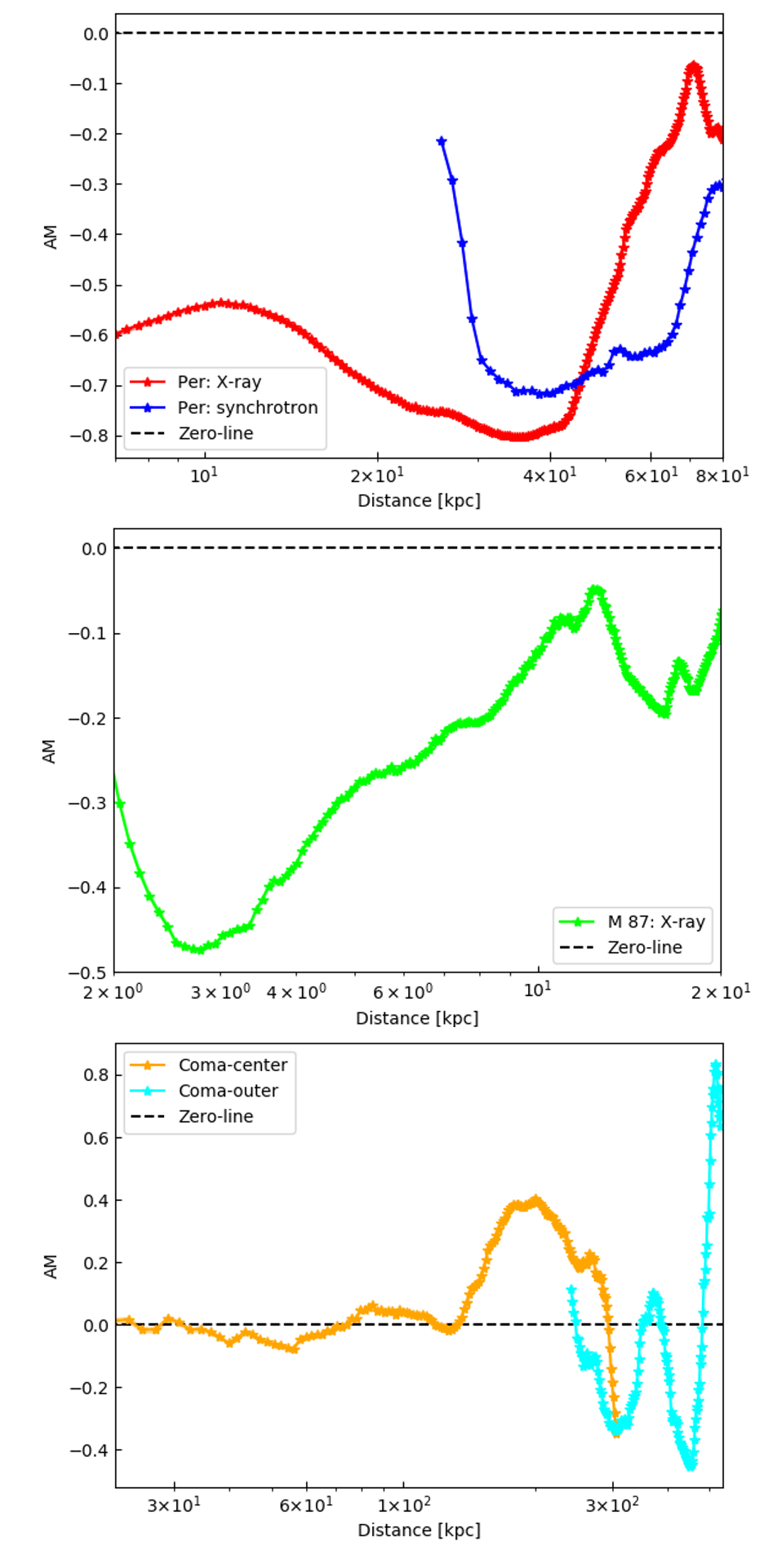}
		\caption{\label{fig:Btang} The relation of AM and the distance away from cluster center. Negative AM implies the magnetic field tends to be tangential, while positive AM means the magnetic field follows the radial direction.}
	\end{figure}

	\section{Discussion}
	\label{sec:diss}
	\subsection{Radial and tangential magnetic fields} 
	
	In this section, we study the orientation of the magnetic fields with respect to the radial direction in all the clusters discussed previously. We employ the AM (see Eq.~\ref{eq.AM} ) to quantify the relative orientation between the magnetic field and the radial direction towards the central black hole. The value of AM is in the range of [-1, 1]. Negative AM implies the magnetic field tends to be tangential, while positive AM means the magnetic field follows the radial direction. The selected regions for the calculation are outlined by dashed circles in Figs.~\ref{fig:perA},\ref{fig:RA},\ref{fig:M87},\ref{fig:A2597}. The coordinates of the central black hole are (49.95$^\circ$, 41.51$^\circ$) and (187.70$^\circ$, 12.39$^\circ$), for Perseus and M 87 in the equatorial coordinate system, respectively.
	
	Fig.~\ref{fig:Btang} shows the relation of AM and the distance away from the cluster center. The AM of Perseus (X-ray) rapidly drops from 0 to $\sim$ -0.6 at distance $r\approx$ 10 kpc and gets to the minimum value $\sim$ -0.8 at $r\approx$ 35 kpc. As for the synchrotron measured Perseus, the drop of AM happens at $r\approx$ 30 kpc, and AM stays $\sim$ -0.7 in the range of [30 kpc, 65 kpc]. 
	Both data sets exhibit the minimum AM at $r\approx$ 35 kpc. 
	In M 87, the AM drops to the minimum value at $r\approx$ 2.5 kpc. The outflow arms contribute to the radial magnetic field at a larger distance, which increases AM, but AM remains negative overall. Note that we remove the AM in the distance corresponding to the minimum scale (20$''$ for Perseus (X-ray), M87, and 1$'$ for Perseus (synchrotron) and Coma) can be resolved by GT. It corresponds to a distance of $\approx$ 7/25 kpc for Perseus (X-ray/synchrotron), $\approx$ 2 kpc for M87 and $\approx$ 27 kpc for Coma away from the center. These values are calculated assuming ideal condition but can be different depending on the noise. For Perseus (synchrotron) and M87, the AM get an increase at a distance smaller than 30 kpc. This likely comes from the artificial effect that GT does not resolve the magnetic field smaller than this distance.
	
	The negative AM in the cores of Perseus and M87 suggest that the magnetic fields are preferentially oriented in the tangential direction. We note that in all three clusters, magnetic draping by rising bubbles may contribute to the tangential component of the magnetic fields. The strongly tangential magnetic fields in Perseus may also be partly attributed to the sloshing arms. However, M87 and Abell 2597 (Fig.~\ref{fig:A2597}) do not have such prominent sloshing arms. Instead, M87 has two outflow arms from AGN feedback, which can randomize magnetic field lines \citep{2016ApJ...818..181Y}. 
	
	For the Coma cluster, we use the centers identified in \citet{IZ19}, i.e., R.A.=12h59m42.67 (J2000) and Dec.=+27$^\circ$56$'$40.9$''$ (J2000) for the Coma-center and R.A.=12h59m22.67 (J2000) and Dec=+27$^\circ$54$'$40.9$''$ (J2000) for the Coma-outer 
	Since Coma is a non-cool-core cluster, the location of its center has a large uncertainty. We test our results by switching the different locations as the cluster center. The trend of AM is similar, but the scale showing minimum AM gets shifted with different centers. In terms of our study, this does not affect the conclusion. Unlike the three cool-core clusters we have analyzed, AM is very close to 0 throughout the core of Coma. 
	At the distance $\sim 100-300$ kpc, our measured positive AM indicates radially oriented magnetic fields which could be associated  with  the  infalling  structures discussed previously. At $r>300$ kpc in Coma, AM shows large fluctuations. Since we have only analyzed a small fraction of the whole annulus, our result is likely biased due to the sampling limit. Future observations and analysis are needed to provide reliable measurements of the mean AM in the outer regions of Coma.

	
	
	\subsection{Tracing the magnetic fields in the ICM with GT}
	As astrophysical flows usually have large Reynolds numbers, the properties of turbulence naturally exhibit everywhere in our universe \citep{1995ApJ...443..209A,2010ApJ...710..853C}. As turbulent eddy is elongating along the direction of local magnetic fields, its density and velocity gradients are preferentially perpendicular to the magnetic fields \citep{GS95,LV99}. As a result, the gradients rotated by 90$^\circ$ are indicating the direction of the magnetic field. This phenomenon has been demonstrated in numerical simulations \citep{2003MNRAS.345..325C,CL02}. Based on this anisotropic property of MHD turbulence, GT is proposed as a new technique for the study of magnetic fields, which has been widely tested for the study of magnetic fields in diffuse ISM and molecular clouds in observation \citep{survey,velac,2019ApJ...874...25G,RNAAS,H2}. As the galaxy cluster also exhibits the properties of MHD turbulence \citep{2019arXiv191106329L,2018PASJ...70...11H,2007PhR...443....1M,2015ApJ...806..103R,2018ApJ...856..162W,IZ19}, we expect that GT is also applicable for the magnetic field studies in ICM. In this work, we present the first prediction of magnetic field morphology in four galaxy clusters, i.e., Coma, Perseus, M 87, and Abell 2597 using X-ray and radio observations. In particular, we find the mean orientation of the magnetic fields in cluster cool cores is preferentially tangential. We also notice that the magnetic fields trace the spiral arm features in the presence of sloshing motion, as predicted by \citet{2011ApJ...743...16Z}, In addition, we find evidence of magnetic draping caused by rising bubbles in cool-core clusters, and in-falling substructures in Coma.
	
	
	Several other methods have been used to study the magnetic fields in galaxy clusters. For example, using the Faraday rotation measurements (RMs), \citet{2016ApJ...823...86A} and \citet{2010A&A...513A..30B} inferred the magnetic field for the radio sources in Coma and  M 87. RM is restricted to radio sources, and cannot be applied to the global magnetic fields in clusters. The advantage of GT is that it can trace the magnetic field on a larger scale. In addition, there is little contribution from foreground in X-ray data or synchrotron emission. Utilizing these data sets, GT therefore directly probes the local magnetic field direction requiring no correction of Faraday rotation. Due to the scale and resolution differences, we do not provide a direct comparison between our study and previous studies using RM. However, we note that both our study and previous RM studies support the existence of turbulent magnetic fields in the ICM. \citet{2010NatPh...6..520P} use polarization measurement to probe magnetic fields around individual galaxies within M 87. They find coherent polarized emission at the leading edges of the moving galaxies, likely caused by magnetic draping. This is consistent with our result showing evidence of magnetic draping around infalling structures in the Coma cluster.

	\subsection{Biases and uncertainties}
	\label{subsec:shock}
	In this section, we discuss possible sources of biases and uncertainties in our analysis. The uncertainties associated with the GT algorithm is presented in Appendix~\ref{appex.d}. 
	
	First, the presence of shocks can potentially introduce problems to the GT. In a strongly magnetized environment, the rapid jump of density at the shock front creates density gradient perpendicular to the shock wave. At the same time, the magnetic field is predominantly perpendicular to the shock front when the relative upstream plasma velocity is greater than the upstream fast wave velocity \citep{shocks}.
	As a result, the intensity gradient becomes parallel to the magnetic field rather than perpendicular, i.e., intensity gradient flips its direction by 90$^\circ$ in front of shocks \citep{YL17b, IGs}. Note that the implementation of ASB can partially suppress the effect from shocks since the shock occupying insufficient pixels will not be resolved. To obtain the actual magnetic field, one should re-rotate the gradient by 90$^\circ$ again in large scale shock front. 
	
	However, if the magnetic field is very weak, the motion of the fluid is essentially equivalent to hydrodynamics. Hydrodynamic shock front then can be parallel to the magnetic field \citep{shocks}. In this case, the intensity gradient in the shock front is still perpendicular to the magnetic field. This phenomenon has been observed by \citet{2010Sci...330..347V} in a radio relic in the outskirts of a galaxy cluster. The ICM is weakly magnetized, and the shock waves in the central regions of galaxy clusters are weak with $M_s\le3$ \citep{2003ApJ...593..599R}. Therefore, we expect the intensity gradient before rotation to be perpendicular to the magnetic field in front of shocks. 
	Further numerical and observational work will help to better understand the effects of shocks in GT analysis of the ICM. 
	
	Also, the theoretical foundation of GT is the anisotropic properties of MHD turbulence, which is a reasonable assumption for small scale structures in clusters. The anisotropic scaling relation is given in \S~\ref{sec:theory}. As for large scales, we expect the weak magnetic fields are constrained by the fluid's motion so that they still follow the anisotropic direction of plasma flow. In this case, we can also use the gradients to trace the magnetic fields. However, once the magnetic field does not follow fluid's motion, our prediction may require correction.

	
	\section{Conclusions}
	\label{sec:conc}
	In this paper, we apply GT to predict the magnetic field morphology in galaxy clusters based on emission maps. We produce the plane-of-the-sky magnetic field maps for four clusters observed in X-rays with {\it Chandra}: Coma, Perseus, M 87, and Abell 2597. We have also applied GT to the synchrotron map of the Perseus core region. We employ the Bayesian analysis to estimate the mean magnetic field orientation in each cluster. Our main findings are summarized as follows:
	\begin{enumerate}
		
		\item We find that
		the magnetic fields follow the sloshing arms in the Perseus cluster, which agrees with the predictions of numerical simulations. In M 87, the magnetic fields follow both arms and shocks. This is not surprising given that the ICM is a high beta, low $M_s$ plasma.
		
		\item The GT-predicted magnetic fields show features typical for magnetic draping. In the central regions of cool-core clusters, a layer of a tangential magnetic field is often seen at the edges of rising bubbles inflated by jets. In Coma, magnetic fields are wrapped around bright, dense halos around NGC4889 and NGC4874 and one of the filaments that are associated with the gas stripped from subclusters merging with the cluster. The latter is consistent with theoretical expectations of magnetic draping.
		

		\item In all three cool-core clusters we have studied, the mean magnetic fields are preferentially oriented tangentially within the cores where temperature gradients are positive. In the isothermal core of Coma, the mean magnetic fields do not show such a preferred direction, while part of the outer region of Coma shows radially oriented mean magnetic fields.
		
		\item There is a broad agreement between the magnetic fields predicted from the X-ray and radio images of the Perseus cluster. 
		Some discrepancies are possibly due to the fact that the X-ray and radio emissions are associated with different   regions within the core.
	\end{enumerate}
	
	Our study is the first to make predictions of the magnetic field orientation on large scales in galaxy clusters. There are potential biases and uncertainties associated with shocks, indirect probes of turbulence, the presence of noise in the data. Further numerical studies, future velocity measurements of the gas in the ICM, and imaging observations with arcsecond resolution and large effective will help us to better understand these uncertainties. Observations using other techniques will also help verify our results.

	\acknowledgments
	We thank John ZuHone for helpful discussions.  A.L. and Y.H. acknowledge the support of the NSF grants AST 1816234 and NASA ATP AAH7546.  The Flatiron Institute is supported by the Simons Foundation.  Y.H., A.L. and I.Z. were supported in part by the National Science Foundation under Grant No. NSF PHY-1748958. I.Z. is partially supported by a Clare Boothe Luce Professorship from the Henry Luce Foundation.
	\software{Julia \citep{2012arXiv1209.5145B}, Paraview \citep{ayachit2015paraview}}
	
	\appendix
	\section{Relation of gradients and structure functions}
	\label{structure}
	
	The GT is directly related to the statistical measures studied for the spectral lines in Lazarian \& Pogosyan (2002, 2004), Kandel et al. (2016, 2017) and for synchrotron emission  and polarization measures studied in Lazarian \& Pogosyan (2012, 2016). Here, following the presentation in Lu et al. (2020) and Lazarian et al. (2020) we briefly provide the relation of the gradients and the structure functions of the observables. 
	
	For a 2D intensity map $I(x,y)=I(\textbf{X})$, the simplest local statistical measurement of its gradient field is the gradient covariance tensor:
	\begin{equation}
	\begin{aligned}
	\sigma_{\nabla_i\nabla_j}&\equiv\langle\nabla_iI(\textbf{X})\nabla_jI(\textbf{X})\rangle=\nabla_i\nabla_jD(\textbf{R})|_{\textbf{R}\to0}\\
	D(\textbf{R})&\equiv\langle(I(\textbf{X}+\textbf{R})-I(\textbf{X})^2)\rangle
	\end{aligned}
	\end{equation}
	which is the zero separation limit of the second derivatives of the field structure function $D(\textbf{R})$. $\langle...\rangle$ denotes the average value. For a statistically isotropic field, the covariance of the gradients is isotropic, i.e., $\sigma_{\nabla_i\nabla_j}=\frac{1}{2}\delta_{ij}\Delta D(\textbf{R})|_{\textbf{R}\to0}$. However, in the presence of the magnetic field, the structure function  becomes orientation dependent, depending on the angle between \textbf{R} and the projected direction of the magnetic field \citep{2012ApJ...747....5L,KLP17b}. This anisotropy is retained in the limit ${\textbf{R}\to0}$ and results in non-vanishing traceless part of the gradient covariance tensor:
	\begin{equation}
	\begin{aligned}
	&\sigma_{\nabla_i\nabla_j}-\frac{1}{2}\sum_{i=x,y}\sigma_{\nabla_i\nabla_i}\\
	&=\frac{1}{2}
	\begin{pmatrix} 
	(\nabla_x^2-\nabla_y^2)D(\textbf{R}) & 2\nabla_x\nabla_yD(\textbf{R})  \\
	2\nabla_x\nabla_yD(\textbf{R}) & (\nabla_y^2-\nabla_x^2)D(\textbf{R}) 
	\end{pmatrix}_{\textbf{R}\to0}\\
	&\ne0
	\end{aligned}
	\end{equation}
	The eigen-direction of the covariance tensor that corresponds to the largest eigenvalue gives the direction of the gradient, which makes an angle $\theta$ with respect to the coordinate x-axis.
	\begin{equation}
	\label{eq.eg}
	\tan\theta=\frac{2\nabla_x\nabla_yD}{\sqrt{((\nabla_x^2-\nabla_y^2)D)^2+(2\nabla_x\nabla_yD)^2}+(\nabla_x^2-\nabla_y^2)D}
	\end{equation}
	\begin{figure*}[p]
		\centering
		\includegraphics[width=0.99\linewidth,height=1.05\linewidth]{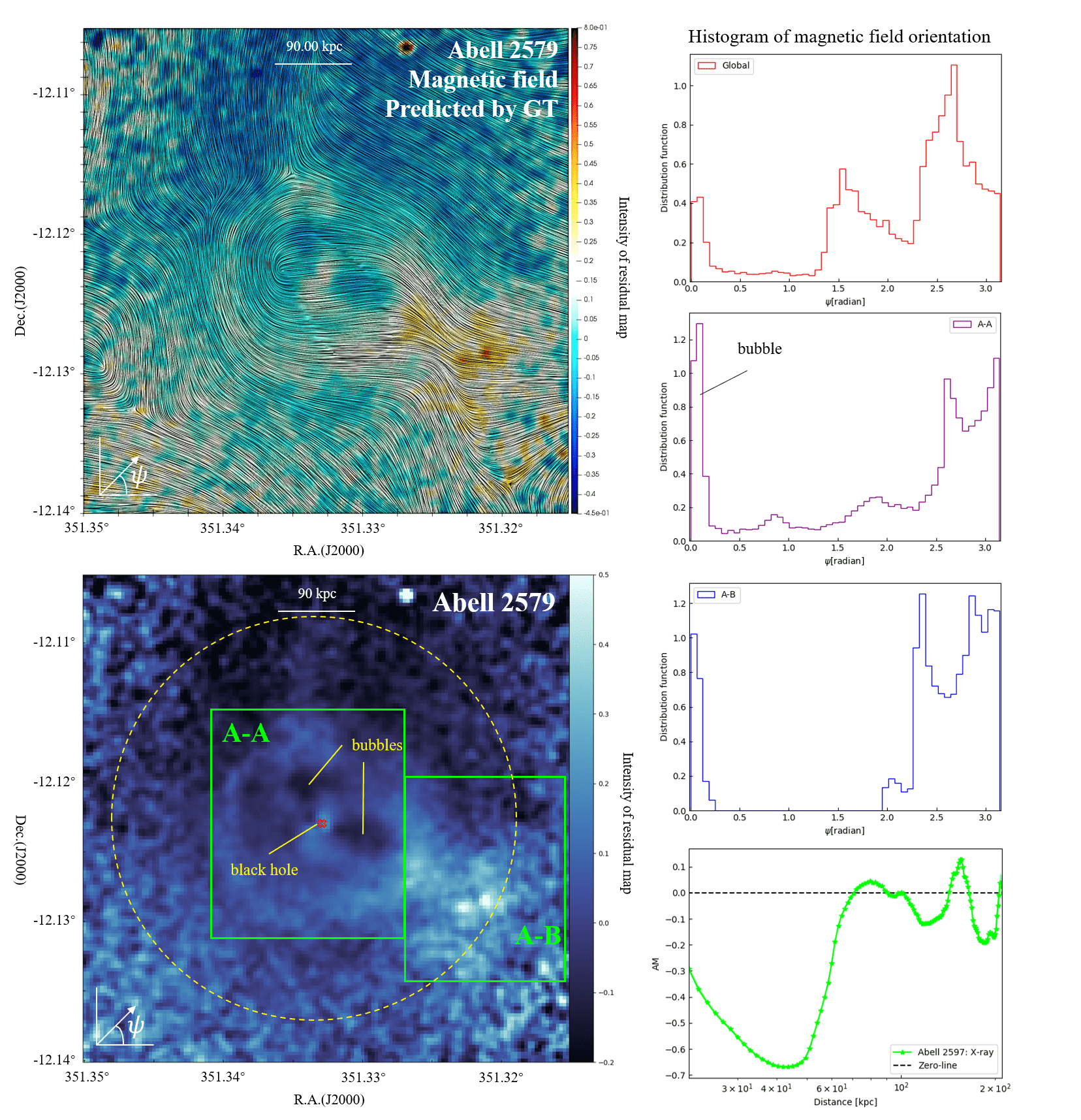}
		\caption{\label{fig:A2597}\textbf{Left top:} the predicted magnetic field morphology of the Abell 2597 cluster from GT. The magnetic field is superimposed in the residual map (i.e., the initial image divided by the best-fitting spherically symmetric $\beta$-model of the surface brightness then minus one) using LIC. \textbf{Left bottom:} the residual image of the Abell 2597 cluster. The cluster is divided into two sub-regions, i.e., A-A and A-B. \textbf{Right:} the histogram of global magnetic field orientation $\psi$ for the Abell 2597 cluster (top, red) and the histograms of magnetic field orientation $\psi$ for the sub-regions A-A (middle, purple) and A-B (middle, blue) respectively. The dashed circles corresponds to $1.35'\approx 120$ kpc, as well as the relation of AM and the distance away from cluster center (bottom).
		}
	\end{figure*}
	
	The anisotropic structure function can be further decomposed in
	angular harmonics \citep{2012ApJ...747....5L, KLP17b}. In Fourier space, the decomposition is over the dependence of the power spectrum $P(\textbf{K})$ on the angle of the 2D wave vector $\textbf{K}$: 
	\begin{equation}
	D(\textbf{R})=-\int d\textbf{K}P(\textbf{K})e^{i\textbf{K}\cdot\textbf{R}}
	\end{equation}
	Denoting the coordinate angle of $\textbf{K}$ as $\theta_K$ and that of the projected magnetic field as $\theta_H$, the power spectrum $P(\textbf{K})$ can be expressed as:
	\begin{equation}
	P(\textbf{K})=\sum_nP_n(K)e^{in(\theta_H-\theta_K)}
	\end{equation}
	we therefore can write derivatives of the structure function as:
	\begin{equation}
	\begin{aligned}
	&\nabla_i\nabla_jD(\textbf{R})=\\
	&\sum_n\int dK K^3P_n(K)\int d\theta_Ke^{in(\theta_H-\theta_K)}e^{iKR\cos(\theta_H-\theta_K)}\hat{K_i}\hat{K_j}
	\end{aligned}
	\end{equation}
	in which $\hat{K_x}=cos(\theta_K)$ and $\hat{K_y}=sin(\theta_K)$. By performing integration over $\theta_K$ , we obtain the traceless anisotropic part:
	\begin{equation}
	\begin{aligned}
	& (\nabla_x^2-\nabla_y^2)D(\textbf{R})=2\pi\sum_ni^ne^{in(\theta-\theta_H)}\times\\
	&\int dKK^3J_n(kR)(P_{n-2}(K)e^{i2\theta_H}+P_{n+2}(K)e^{-i2\theta_H})\\
	& \nabla_x\nabla_yD(\textbf{R})=\pi\sum_ni^{n+1}e^{in(\theta-\theta_H)}\times\\
	&\int dKK^3J_n(kR)(-P_{n-2}(K)e^{i2\theta_H}+P_{n+2}(K)e^{-i2\theta_H})\\
	\end{aligned}
	\end{equation}
	in which $J_n(kR)$ is the Bessel function. In the limit $\textbf{R}\to0$, only $n$ = 0 term survives and we have:
	\begin{equation}
	\begin{aligned}
	\label{eq.D}
	(\nabla_x^2-\nabla_y^2)D(\textbf{R})&=[2\pi\int dKK^3P_2(K)]\cos2\theta_H\\
	2\nabla_x\nabla_yD(\textbf{R})&=[2\pi\int dKK^3P_2(K)]\sin2\theta_H
	\end{aligned}
	\end{equation}
	By substituting this result into Eq.~\ref{eq.eg}, we find
	that the eigendirection of the gradient variance has the form:
	\begin{equation}
	\tan\theta=\frac{A\sin2\theta_H}{|A|+A\cos2\theta_H}=
	\begin{cases}
	\tan\theta_H, A>1\\
	-\cot\theta_H, A<1\\
	\end{cases}
	\end{equation}
	where $A=2\pi\int dKK^3P_2(K)$. The sign of $A$ depends on the spectral quadrupole $P_2(K)$. As it is discussed in Lazarian et al. (2020) the quadrupole is negative for Alfv\'{e}n and slow modes. In contrast, fast modes in low-$\beta$ plasma produce positive quadrupole. Therefore, the intensity gradients are perpendicular to the projected magnetic field for Alfv\'{e}n and slow modes, while parallel to the projected magnetic field for low-$\beta$ fast-modes.
	\begin{figure*}[t]
		\centering
		\includegraphics[width=.99\linewidth,height=0.83\linewidth]{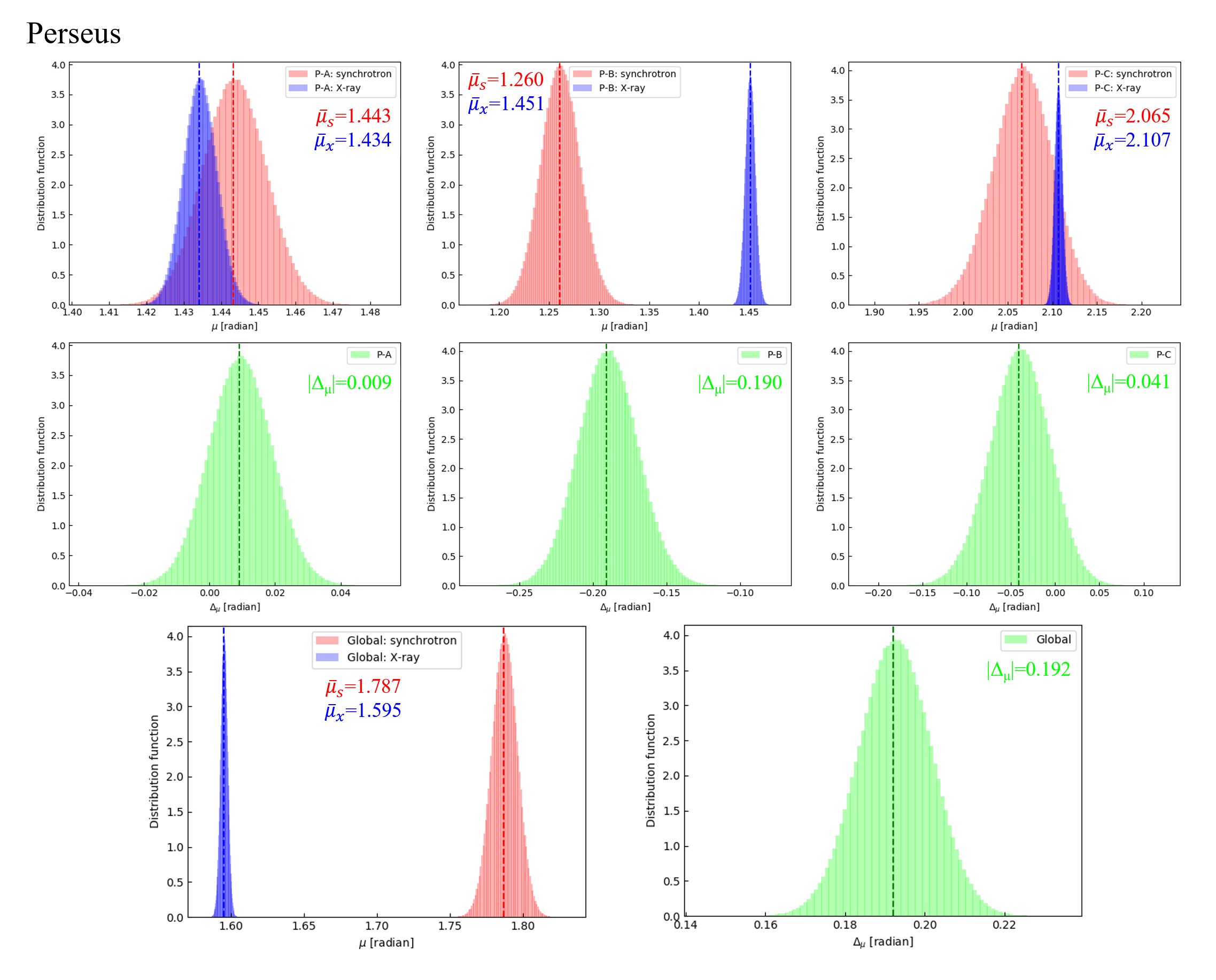}
		\caption{\label{fig:mu}\textbf{Top:} the distribution of the mean magnetic field predicted from synchrotron ($\bar{\mu}_s$, red) and X-ray data ($\bar{\mu}_x$, blue). The dashed lines are indicating the median value. The distribution is drawn for each sub-region P-A, P-B, and P-C respectively using the Bayesian analysis (see Fig.~\ref{fig:perA} and Fig.~\ref{fig:RA} for details of each sub-region). \textbf{Bottom:} the distribution of the relative mean angle $\Delta_\mu=\bar{\mu}_s-\bar{\mu}_x$ of the magnetic field predicted from synchrotron and X-ray data.}
	\end{figure*}
	\begin{figure*}[t]
		\centering
		\includegraphics[width=1.0\linewidth,height=0.7\linewidth]{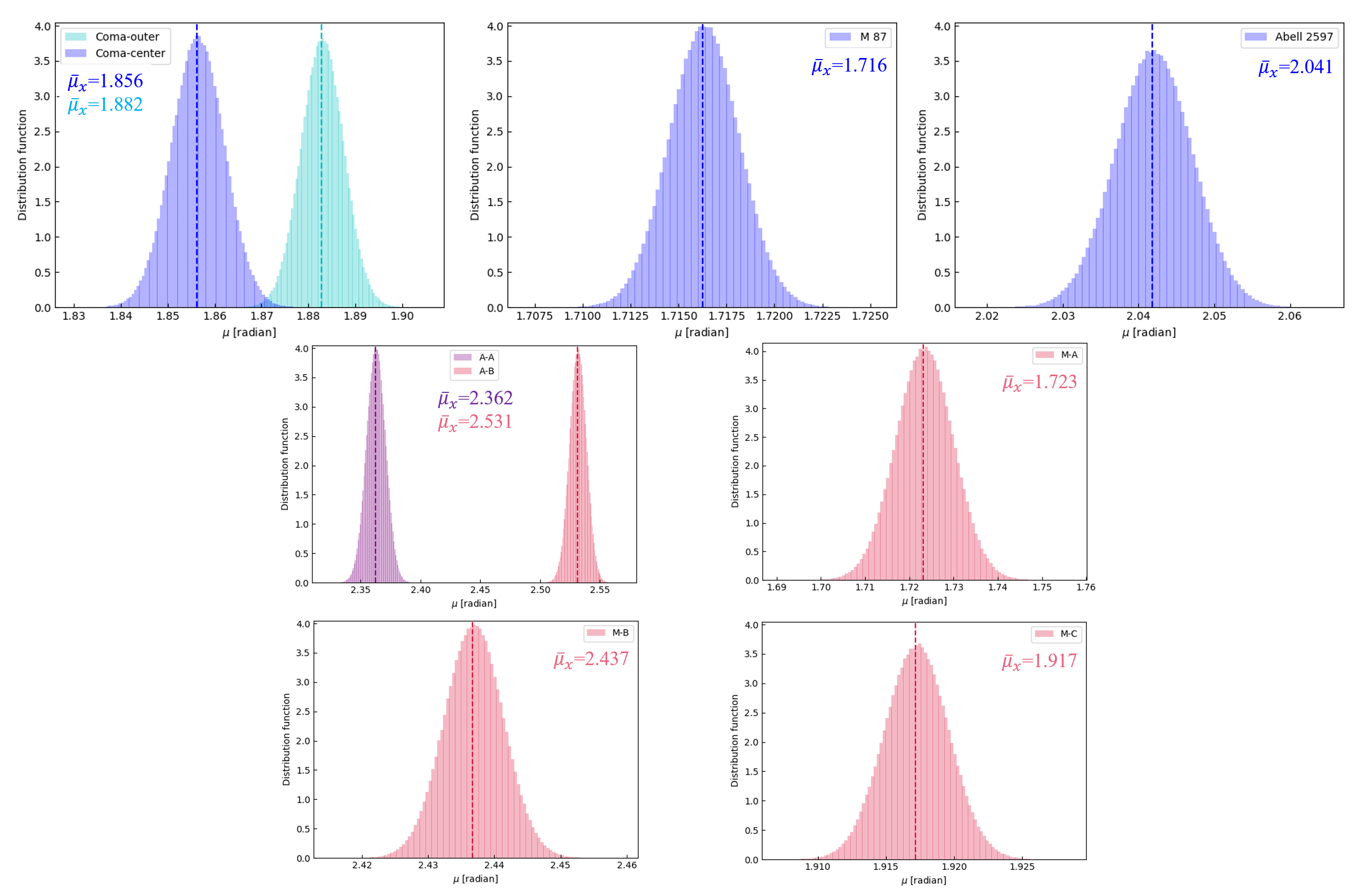}
		\caption{\label{fig:sub-hist}\textbf{Top:} the distribution of the global mean magnetic field $\bar{\mu}_x$ predicted from X-ray data. The dashed lines are indicating the median value. The distribution is drawn for Coma-outer and Coma-center (left), M87 (middle), and Abell 2597 (right) respectively using the Bayesian analysis. \textbf{Bottom:} the distribution of the mean magnetic field $\bar{\mu}_x$ for sub-regions A-A and A-B (upper left), M-A (upper right), M-B (bottom left), and M-C (bottom right), see Fig.~\ref{fig:M87} and Fig.~\ref{fig:A2597} for details of each sub-region.}
	\end{figure*}
	
	\section{Predicted magnetic field morphology in Abell 2579 cluster}
	\label{ap.A}
	
	The recipe of GT is also repeated for the cluster Abell 2597 using X-ray residual maps. 
	The corresponding residual map and predicted magnetic field are shown in Fig.~\ref{fig:A2597}. The bubbles exhibit distinct magnetic field morphology. In the histograms, we see the magnetic field in A-A locates in the angle range $\simeq 2.5\sim\pi$ and $0\sim0.2$ (note gradient does not distinguish 0 and $\pi$), as well as the magnetic field in the A-B sub-region. 
	To sum up, in all four clusters, we find the magnetic field predominately follows the outflow structures. This can be easily interpreted. The clusters are typically in super-Alfv\'{e}n environment, which means the dominance of the magnetic field is relatively weaker than turbulence. In this case, turbulence will alter the magnetic field geometry so that it tends to align parallel the density structures because of the magnetic flux freezing condition. As for the calculation of the tangential magnetic field, we use the coordinate of the central black hole (351.333$^\circ$, -12.123$^\circ$) in the equatorial coordinate system. The minimum AM of the Abell 2597 cluster appears at $r\approx$ 40 kpc. We can expect the scale of the most significantly tangential magnetic field is $\approx40$ kpc for Abell 2597. 
	\section{Bayesian analysis}
	\label{appendx}
	
	To estimate the mean magnetic field angle predicted from the synchrotron and X-ray data set, we employ the Bayesian t-test analysis in this section. The Bayesian analysis is based on Bayes’ theorem to obtain the conditional distribution for the unobserved quantities given the data $x$ which is known as the posterior distribution $p(\mu|x)$: 
	\begin{equation}
	\label{eq:13}
	p(\mu|x)=\frac{p(\mu)p(x|\mu)}{\int p(\mu)p(x|\mu)d\mu}\propto p(\mu)p(x|\mu)
	\end{equation}
	where $\mu$ is a the mean magnetic field angle we would like to know, $p(x|\mu)$ is the probability of getting $x$ given $\mu$, which is called the likelihood. $p(\mu)$ is the prior probability of $\mu$ before considering the data. Eq.~\ref{eq:13} can be understood as follow: prior to observing data, one assigns a prior belief $p(\mu)$ to $\mu$. Once the data x have been observed, one updates this prior belief to a posterior belief $p(\mu|x)$ by multiplying the prior $p(\mu)$ by the likelihood $p(x|\mu)$ \citep{2018arXiv181203092F}.
	\begin{figure*}[t]
		\centering
		\includegraphics[width=0.81\linewidth,height=1.27\linewidth]{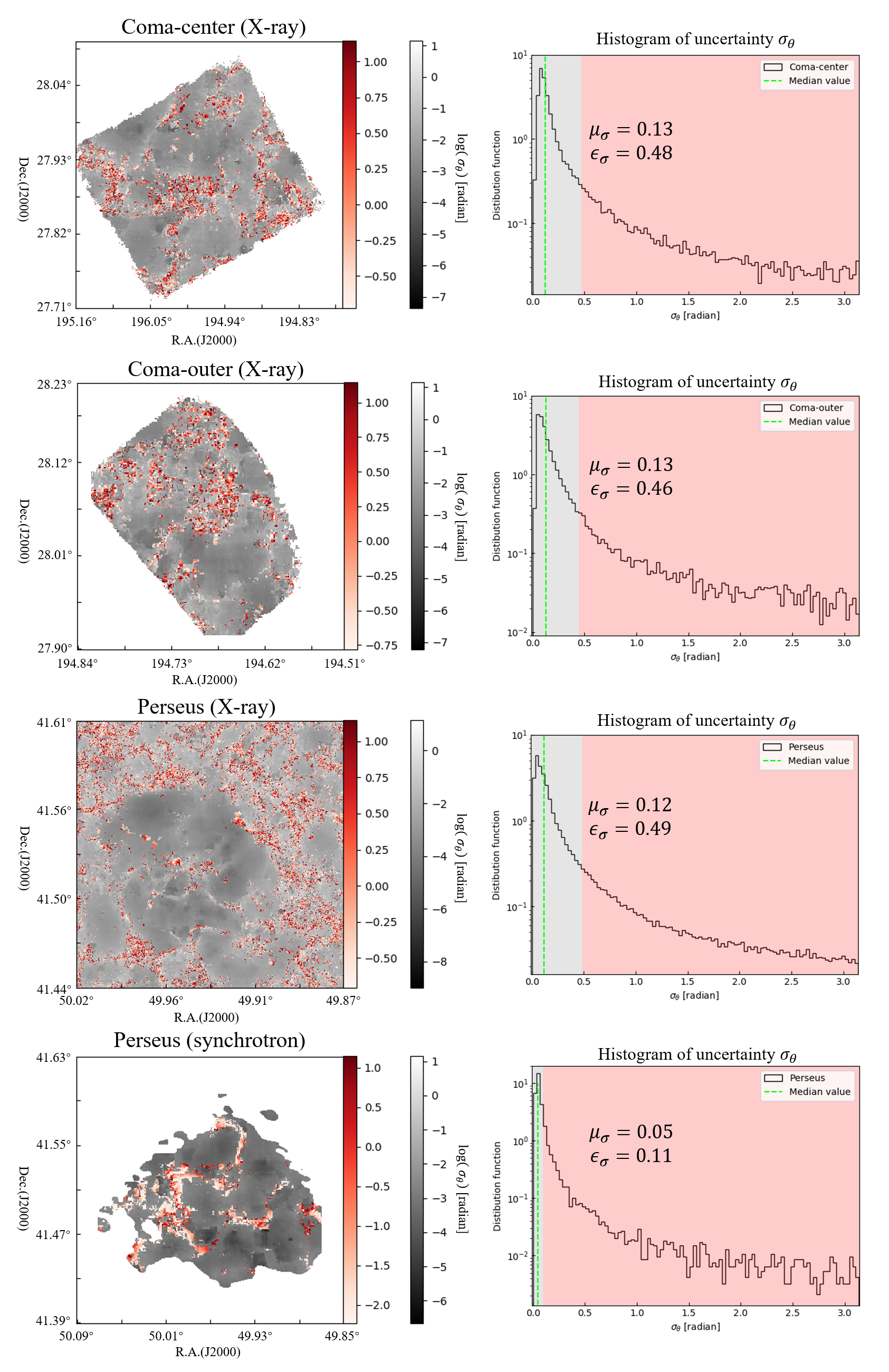}
		\caption{\label{fig:error}\textbf{Left:} The logged uncertainty maps of the predicted magnetic field morphology for clusters Coma and Perseus. The uncertainty $\sigma_\theta$ is in the range [0,$\pi$). High uncertainty pixels, in which the uncertainty is larger than 1$\sigma$ level of the global system, are highlighted by red color. Grey color represents the low uncertainty pixels. \textbf{Right:} the histogram of uncertainty $\sigma_\theta$ for the four clusters. $\mu_\sigma$ is the median value of $\sigma_\theta$ and $\epsilon_\sigma$ represents the uncertainty value in 1$\sigma$ level. We use red background to indicate the range in which $\sigma_\theta$ is larger than $\epsilon_\sigma$.}
	\end{figure*}
	\begin{figure*}[t]
		\centering
		\includegraphics[width=0.90\linewidth,height=0.73\linewidth]{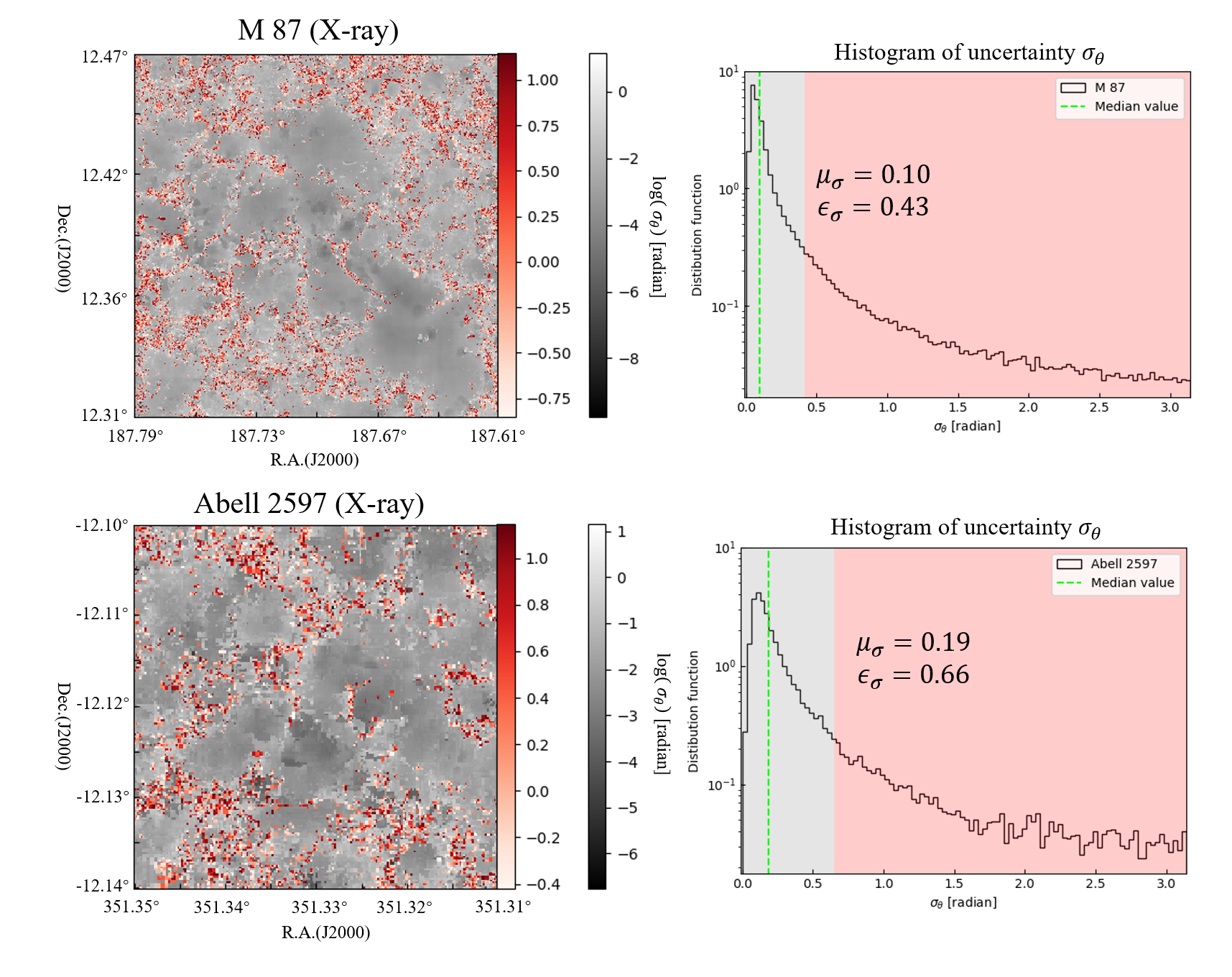}
		\caption{\label{fig:error2}\textbf{Left:} The logged uncertainty maps of the predicted magnetic field morphology for the clusters M 87 and Abell 2597. The uncertainty $\sigma_\theta$ is in the range [0,$\pi$). High uncertainty pixels, in which the uncertainty is larger than 1$\sigma$ level of the global system, are highlighted by red color. Grey color represents the low uncertainty pixels. \textbf{Right:} the histogram of uncertainty $\sigma_\theta$ for the four clusters. $\mu_\sigma$ is the median value of $\sigma_\theta$ and $\epsilon_\sigma$ represents the uncertainty value in 1$\sigma$ level. We use red background to indicate the range in which $\sigma_\theta$ is larger than $\epsilon_\sigma$.}
	\end{figure*}
	
	The first step in a Bayesian analysis is to specify the probability model of the likelihood and the prior distribution. The T distribution is usually considered as a robust choice to the analysis, as it is less sensitive to outlier observations and requires no information on the quantity's standard deviation \citep{Kruschke2013}. $p(x|\mu)$ then can be expressed as:
	\begin{equation}
	p(x|\mu)=\frac{\Gamma(\frac{\nu+1}{2})}{\Gamma(\frac{\nu}{2})}[1+\frac{(x-\mu)^2}{\nu}]^{-\frac{\nu+1}{2}}
	\end{equation}
	where $\nu$ is the degrees-of-freedom chosen as the size of the sample minus one and $\Gamma$ represents gamma function. As for $p(\mu)$, we apply a normal prior $\mathcal{N}$ on it based on the central limit theorem, which says the sampling distribution of a statistic will follow a normal distribution, as long as the sample size is sufficiently large:
	\begin{equation}
	p(\mu)=\mathcal{N}(\langle x\rangle, \sigma_x)
	\end{equation}
	here $\langle x\rangle$ and $\sigma_x$ are the average value and the standard deviation of x respectively. The output mean magnetic field $\bar{\mu}$, which follows the distribution $p(\mu|x)$, is obtained through the Markov chain Monte Carlo (MCMC) sampling method \citep{vanRavenzwaaij2018}. In this work, we consider $1\times10^5$ samples from MCMC, which gives sufficiently statistical information. 
	
	As the given data $\psi$ follows angular statistics, we should modify the Bayesian analysis accordingly. We firstly apply the Bayesian analysis to both sine and cosine components of $\psi$, producing two outputs $\overline{\cos(\mu)}$ and $\overline{\sin({\mu})}$. The desired $\bar{\mu}$ is calculated from $\bar{\mu}=\tan^{-1}(\overline{\sin({\mu})}/\overline{\cos(\mu)})$.
	
	\subsection{Comparison of the magnetic field predicted from X-ray and synchrotron data}
	
	Fig.~\ref{fig:mu} presents the distribution of the mean magnetic field $\bar{\mu}$ obtained from the Bayesian analysis for the three sub-regions, P-A, P-B, and P-C. We only consider the areas where the X-ray and synchrotron data are overlapping. We denote the value of $\bar{\mu}$ calculated from the synchrotron data as $\bar{\mu}_s$, while $\bar{\mu}_x$ is for the X-ray data. For the sub-region P-A, the median values $\bar{\mu}_s=1.443$ and $\bar{\mu}_x=1.434$. The distribution of $\bar{\mu}_x$ is narrower than the one of $\bar{\mu}_s$ since the X-ray data gives more initial samples $\psi$. The difference between $\bar{\mu}_s$ and $\bar{\mu}_x$ is calculated as $\Delta_\mu=\bar{\mu}_s-\bar{\mu}_x$. The distribution of $\Delta_\mu$ spreads from $\sim-0.02$ to $\sim-0.04$ with an absolute median value of $|\Delta_\mu|=0.09$. As for P-B, the distributions of $\bar{\mu}_s$ and $\bar{\mu}_x$ are not overlapped giving $\bar{\mu}_s=1.260$ and $\bar{\mu}_x=1.451$. The absolute median value of $|\Delta_\mu|$ is 0.190. Similarly for P-C, we have median values $\bar{\mu}_s=2.065$ and $\bar{\mu}_x=2.107$ with $|\Delta_\mu|=0.041$. Also, we repeat the Bayesian analysis for the global overlapped areas for the X-ray and synchrotron data. The results are presented in Fig.~\ref{fig:mu}. We get $\bar{\mu}_s=1.787$, $\bar{\mu}_x=1.595$, and $|\Delta_\mu|=0.192$.
	
	\section{Mean magnetic field direction}
	The Bayesian analysis is also applied to the clusters Coma, M87, and Abell 2597. As shown in Fig.~\ref{fig:sub-hist}, the global mean magnetic field direction $\bar{\mu}_x$ of Coma-center is 1.856 ($\approx106.34^{\circ}$) and  $\bar{\mu}_x=1.882$ ($\approx107.83^{\circ}$). As for the M 87 and Abell 2597 clusters, we have the global $\bar{\mu}_x=1.716$ ($\approx98.31^{\circ}$) and $\bar{\mu}_x=2.041$ ($\approx116.94^{\circ}$). In Fig.~\ref{fig:sub-hist}, we also plot the the distributions of $\bar{\mu}_x$ for the sub-regions A-A and A-B of Abell 2597, as well as M-A, M-B, and M-C of M 87 (see Fig.~\ref{fig:M87} and Fig.~\ref{fig:A2597} for details of each sub-region). As a result, we get $\bar{\mu}_x=2.362$ for A-A and $\bar{\mu}_x=2.362$ for A-B. While we have $\bar{\mu}_x=1.723$ for M-A, $\bar{\mu}_x=2.437$ for M-B, and $\bar{\mu}_x=1.917$ for M-C.
	
	\subsection{Uncertainty of the predicted magnetic field direction}
	\label{appex.d}
	The two significant uncertainties of the predicted magnetic field can come from the systematic error in the observation map and the ASB algorithm. Recall that the ASB takes a sub-region and fits a corresponding Gaussian histogram of gradient's orientation. The output of ASB only takes the statistically most crucial angle, i.e., the angle of orientation corresponding to the Gaussian fitting peak value of the histogram. This procedure incidentally suppresses the part of the systematic noise in the observation map. The uncertainty of the ABS its own in every single pixel can be considered as the error $\sigma_\psi$ from the Gaussian fitting algorithm within 95\% confidence level. 
	
	In Fig.~\ref{fig:error} and Fig.~\ref{fig:error2}, we plot the uncertainty of the predicted magnetic field direction for the four clusters. The uncertainty is expressed in log scale, i.e., $\log(\sigma_\psi)$ in which $\sigma_\psi$ is in the range [0, $\pi$). We also highlight two important values: the median value $\mu_\sigma$ and the value $\epsilon_\sigma$ corresponding to 1$\sigma$ level. We use red color to distinguish the data points whose uncertainty is larger than $\epsilon_\sigma$ and grey color is indicating the uncertainty less than $\epsilon_\sigma$. In the case that the uncertainty is maximum, i.e., $\sigma_\psi\approx\pi$, it indicates the systematic noise in the observation map is extremely large and ASB does not produce the appropriate measurement. The solution is proposed in \S~\ref{subsub:RHT}, i.e., the rotational histogram test. After rotating the residual map by $90^\circ$, the ABS results in a similar angle of $\psi$. The corresponding mask and interpolation are then applied to those noisy pixels. We present the original map of uncertainty in Fig.~\ref{fig:error} for the reminding of interpolated gradients.
	
	To estimate the global uncertainty in our prediction, we plot the histograms of $\sigma_\psi$ in Fig.~\ref{fig:error}. The histograms exhibit a median value $\mu_\sigma =$ in the range from 0.10 to 0.19, corresponding to $5.72^\circ \sim 10.88^\circ$ for the X-ray measured clusters Coma, Perseus, M 87, and Abell 2597. As for the synchrotron measured Perseus cluster, we have $\mu_\sigma =0.05\approx2.86^\circ$. The uncertainty of $\mu_\sigma$ can be given by the standard error of the mean, which is very insignificant due to the large sample size. Note the $\mu_\sigma$ is calculated from the raw uncertainty map, i.e., without the implementation of the pseudo-Stokes parameters and the rotational histogram test, which help with reducing noise and uncertainty. $\mu_\sigma$ here therefore should be the extreme value.

	%
	
	\vspace{5mm}
	

	


	
	
\end{document}